\documentstyle[twocolumn,epsf]{article}

\def\thebibliography#1{\section*{\normalsize \bf References 
 }\list
 {[\arabic{enumi}]}
  {\settowidth\labelwidth{[#1]}\leftmargin\labelwidth
 \advance\leftmargin\labelsep
 \usecounter{enumi}}
 \def\newblock{\hskip .11em plus .33em minus .07em}
 \sloppy\clubpenalty4000\widowpenalty4000
 \sfcode`\.=1000\relax}

\begin{document}

\twocolumn[

\begin{center} \large \bf
  Weak-coupling approach to the semi-infinite Hubbard model: \\
  Non-locality of the self-energy
\end{center}
\vspace{-3mm}

\begin{center} 
   M. Potthoff and W. Nolting
\end{center}
\vspace{-6mm}

\begin{center} \small \it 
   Lehrstuhl Festk\"orpertheorie,
   Institut f\"ur Physik, 
   Humboldt-Universit\"at zu Berlin, 
   Germany
\end{center}
\vspace{2mm}

\begin{center}
\parbox{141mm}{
The Hubbard model on a semi-infinite three-dimensional lattice
is considered to investigate electron-correlation effects at 
single-crystal surfaces. The standard second-order perturbation 
theory in the interaction $U$ is used to calculate the electronic 
self-energy and the quasi-particle density of states (QDOS) in the 
bulk as well as in the vicinity of the surface. Within a real-space 
representation we fully account for the non-locality of the 
self-energy and examine the quality of the local approximation. 
Numerical results are presented and discussed for the three 
different low-index surfaces of the simple-cubic lattice. 
Compared with the bulk significant differences can be found for 
the top-layer local self-energy, the imaginary part of which is 
energetically narrowed and has a reduced total 
weight. The non-local parts of the self-energy $\Sigma_{ij}(E)$ 
decrease with increasing distance between the sites $i$ and $j$. 
At the surface and for the three-dimensional bulk their decrease 
is faster than for a two-dimensional lattice.
For all surfaces considered the effects of the non-local parts of 
the self-energy on the QDOS are found to be qualitatively the same 
as for the bulk: The weight of the quasi-particle 
resonance at the Fermi 
energy is lowered while the high-energy charge-excitation peaks 
become more pronounced.
The main structures in the layer-dependent spectra are already 
recovered within the local approximation; taking into account 
the nearest-neighbor non-local parts turns out to be an excellent
approximation. Due to the reduced coordination number for sites 
at the very surface, 
the top-layer QDOS is narrowed. Contrary to the the free ($U=0$)
system, quasi-particle damping results in a comparatively weak
layer dependence of the QDOS generally. Pronounced surface effects
that are related to surface Friedel oscillations show up as a
fine structure within the resonance around the Fermi level.

\vspace{4mm} 
{\bf PACS:} 71.27.+a, 73.20.At, 75.10.Lp
}
\end{center}
\vspace{8mm} 
]

{\center \bf \noindent I. INTRODUCTION \\ \mbox{} \\} 

The Hubbard model \cite{Hub63,Gut63,Kan63} is one of the most 
extensively studied models for interacting fermions on a lattice. 
While originally it was set up to study the conditions for 
metal-insulator (Mott) transitions and 
spontaneous magnetism, it nowadays quite generally contributes 
to the basic understanding of different types of prominent 
correlation effects.

The mutual influence of electron correlations and the geometry of 
the underlying lattice is one of the central aspects of recent 
research. The simplest example for the decisive role of the 
geometry is given by the lattice dimension $D$. 
In the extreme cases $D=1$ \cite{LW68,d1one,d1two} and $D=\infty$ 
\cite{MV89,Vol93,GKKR96} 
the Hubbard model exhibits completely different physical
properties. Also for a given dimension, say $D=3$, the lattice
structure is important. The exact results by Nagaoka \cite{Nag66}
as well as approximate treatments \cite{NB89,BdKNB90,HN97a}
represent illustrating examples for the interplay between
magnetic order and the lattice type. 

The limit $D=\infty$ is of special interest since here the effects
of the lattice geometry are suppressed as much as possible while 
the model itself remains non-trivial. The on-site Green function 
$G_{ii\sigma}(E)$ depends on the geometrical structure merely via 
the free Bloch-density of states (BDOS), and the self-energy 
$\Sigma_{\sigma}(E)$ becomes $\bf k$-independent or site-diagonal
\cite{MH89a}. However, even for $D=\infty$ the dependence on the 
lattice type, i.~e.\ on the BDOS is important as has been proved,
for example, by
the recent investigations on ferromagnetism for the hypercubic and
an fcc-type lattice \cite{VBH+97,OPK97}.

It has been argued \cite{Vol93} that the essential physical 
properties of the $D=\infty$ Hubbard model are comparable to
those for finite dimensions. A solution of the $D=\infty$ Hubbard 
model would thus provide for a proper mean-field theory. 
Actually this argument can be considered as a justification of 
the local approximation for the self-energy, i.~e.\ the neglect 
of its $\bf k$-dependence or its non-local parts in the case
of finite dimension. Since the contribution of non-local parts 
of the self-energy to relevant physical quantities scales with 
$1/D$ \cite{Vol93}, the local approximation becomes less
accurate in the opposite limit $D\mapsto 1$. Starting from the
local approximation and successively including nearest-neighbor,
next-nearest neighbor and further correction terms, corresponds 
to an expansion in $1/D$. An instructive demonstration of this 
fact has been worked out by Schweitzer and Czycholl \cite{SC91} 
for $D=1-3$. The expansion converges even for $D=1$; for $D=2$ 
all corrections beyond the 5-th neighbor shell are found to be 
negligibly small, and for $D=3$ the local approximation is quite 
appropriate.

The interplay between geometric and electronic structure becomes 
more important and interesting when the translational symmetry of 
the lattice is incomplete. This situation is realized e.~g.\ in 
the case of a film of finite thickness composed of two-dimensional 
layers or at the surface of a three-dimensional lattice. Our study 
aims at the latter case, the semi-infinite Hubbard model. The 
semi-infinite Hubbard model may be interesting for different
reasons. First of all, it comes closer to the experimental 
situation. Because of the finite inelastic mean free path of 
low-energy electrons in the solid \cite{Pow88}, surface effects are 
often non-negligible or even dominating in electron spectroscopy.
Furthermore, surface effects may be interesting of their own.
Particularly, the semi-infinite Hubbard model can contribute to 
the understanding of phenomena related to surface 
magnetism. The combined influence of electron-correlation as well 
as of surface effects is important for the study of magnetic surface 
states, of magnetic moments at the surface and a possibly enhanced 
surface Curie temperature etc. To our 
knowledge little effort has been spent on these topics up 
to now; the investigation of itinerant electron models with reduced
translational symmetry (thin films, surfaces) is still at the
very beginning (see Refs.\ \cite{PM95,PN95,PN96,PN97b,PN97a}). 
On the other hand, there has been extensive work on the 
semi-infinite Ising and Heisenberg model \cite{isihei}. These
localized spin models, however, have to be considered as effective
models where the magnetic order is more or less pretexted rather
than derived consistently from the electronic structure.

Despite its conceptual simplicity, an exact solution of the 
Hubbard model or an approximation scheme that is reliable in 
the entire parameter space is not available for $D=3$ up to 
now. The presence of the surface and the corresponding breakdown
of translational symmetry in the surface normal direction still
complicates the problem, and thus approximations must be tolerated.
One of the main purposes of the present paper is to contribute to a 
clarification of the status of the local approximation for the 
self-energy in the context of the semi-infinite Hubbard model. 
This promises to be an interesting question since a priori it is
not obvious whether a solid surface in this respect should be 
regarded as being close to $D=2$ or close to $D=3$. However, while 
the local approximation may be sufficient in the case $D=3$, 
it is less accurate 
in $D=2$ \cite{SC91,MEHM95,BE95}. It can be expected that the degree
of non-locality of the self-energy will be tightly connected with 
the reduction of the coordination number of sites at the very 
surface. One should try to substantiate a reasoning based 
on coordination numbers by means of a more quantitative analysis.
This includes the investigation of different (low-index) surfaces
of the same lattice.
While non-local parts $\Sigma_{ij\sigma}(E)$ for $i\ne j$ are 
expected to decrease for distances 
$||{\bf R}_i - {\bf R}_j|| \mapsto \infty$,
the number of atoms within a shell of increasing radius $R$ 
around the site $i$ may increase. This raises the question of the 
effect of non-local contributions on the quasi-particle density of 
states and the changes near the surface.

In the present paper we try to tackle the mentioned problems 
within the weak-coupling approach to the semi-infinite Hubbard
model, i.~e. within second-order perturbation theory (SOPT) in 
the interaction $U$. For the translationally symmetric Hubbard 
model this is a standard approximation scheme 
\cite{SC91,CKdCL78,Kle79,TDS80a,Kis87,TT88,MH89b,BJ90,SAS92,SSHZ95}.
Its numerical evaluation in $\bf k$-space includes non-trivial
multiple integrals. An efficient real-space algorithm for the 
calculation of the self-energy
$\Sigma_{ij\sigma}(E)$ was introduced in Refs.\ \cite{SC91,SC90} 
and is easily seen to be applicable also in the case 
of reduced translational symmetry. The SOPT will yield exact 
results for infinitesimal $U=0^+$ and will give first hints at 
least for finite $U$. An extrapolation of the results to strong 
interaction $U$ (compared with the Bloch-band width $W$) must be 
questioned, of course. On the other hand, the 
comparison of the SOPT results for moderate interaction $U$ 
\cite{BJ90} with the results of a strong-coupling moment method 
\cite{NB89} is encouraging since qualitatively both approaches 
agree well. In Ref.\ \cite{BJ90} Bulk and Jelitto employed the
SOPT around the Hartree-Fock (HF) solution. This approach turns out 
to be superior compared with the plain SOPT in which free instead of
Hartree-Fock propagators are used in the calculation of the proper
irreducible self-energy diagram. Neither plain SOPT nor SOPT-HF,
however, is a conserving approximation in the sense of Baym and 
Kadanoff \cite{BK61}. Contrary, the self-consistent SOPT (using
the fully renormalized propagators) does represent a 
$\Phi$-derivable theory \cite{MH89b,MMH91}.

Most of the results presented here have been calculated for 
$U=0^+$. In this case the different versions of the weak-coupling 
theory become identical. Considering an infinitesimal interaction,
however, does not allow to study the manifestation of correlation 
effects in the quasi-particle density of states. Therefore, we
will also consider finite interaction $U$, but restrict ourselves
to the symmetric situation of a paramagnet at half-filling 
($n \equiv 2 \langle n_{i\sigma} \rangle = 1$) in this case. 
The symmetric case is somehow exceptional; the following well-known
arguments substantiate that SOPT-HF may yield qualitatively
correct results even for moderate interaction $U\sim W$ at $n=1$. 
Firstly, SOPT-HF exactly reproduces the atomic limit ($W=0$) at 
half-filling. Secondly, the energetic positions as well as the
spectral weights of both high-energy charge-excitation peaks that
show up in the spectrum for $U\gg W$ can be calculated 
exactly within a strong-coupling perturbation approach (expansion 
in $W/U$) \cite{HL67,EOMS94}. For the case of half-filling SOPT-HF 
can easily be shown to reproduce these exact results. We thus
believe that the method reasonably interpolates between the 
weak- ($U\ll W$) and the strong-coupling ($U\gg W$ and $W=0$) 
regimes. 
On the contrary, the self-consistent 
SOPT fails with respect to all points mentioned. In our opinion 
this cannot be outweighted by its advantages, namely to represent 
a conserving theory and thus to respect the Luttinger sum rule 
\cite{LW60}. Since the SOPT-HF scheme preserves particle-hole 
symmetry \cite{Mer77}, the sum rule is fulfilled anyway in the case 
of half-filling. 
The last argument regards the high-energy behavior of 
the SOPT-HF self-energy. At half-filling the first two coefficients 
(for the local part three coefficients) within an $1/E$ expansion 
turn out to be exact (see Ref.\ \cite{KK96,PWN97}). This also 
implies that the first three moments of the quasi-particle density 
of states can be reproduced. 
Another reason for using SOPT-HF is a pragmatic 
one: For the general (non-symmetric) case, self-consistency
with respect to the occupation number $\langle n_{i\sigma} \rangle$
and the chemical potential $\mu$ is required. At half-filling this 
is achieved trivially, we have $\mu=U/2$ (this also holds for the 
semi-infinite lattice). 
Since our main interest is focussed on the validity of the local
approximation for the self-energy, we may set aside the fact that
the Hubbard model at half-filling is expected to be an 
antiferromagnet \cite{and63}. Throughout the paper we exclusively
consider the (possibly metastable) paramagnetic phase.

The next section briefly introduces the theoretical concept. In
section III we present and discuss the results. All calculations 
have been performed for zero temperature $T=0$. We have limited
our investigations to the simple-cubic (sc) lattice. In most 
cases the (100) surface was considered, but also the
more open sc(110) and sc(111) surfaces were taken into account 
for comparison. A summary and the conclusions are given 
in section IV.

{\center \bf \noindent II. THEORY AND CALCULATIONS \\ \mbox{} \\} 

We consider interacting electrons moving through a semi-infinite
lattice. The dynamics of the system is described by the Hubbard 
model. Using standard notations the Hamiltonian reads:

\begin{equation}
  H = \sum_{\langle ij \rangle \sigma} 
  \left( T_{ij} - \mu \delta_{ij} \right)
  c^\dagger_{i\sigma} c_{j\sigma} 
  + \frac{1}{2} U \sum_{i \sigma} n_{i\sigma} n_{i -\sigma} \: .
\label{eq:hubbard}
\end{equation}
The system geometry is implicit: $c^\dagger$ ($c$) is the creation 
(annihilation) operator for a valence electron. The lower indices
specify the orbital of the one-particle basis. $\sigma=\uparrow , 
\downarrow$ accounts for the spin direction, and $i$ and $j$ 
refer to the sites within the semi-infinite lattice. We consider 
the system to be built up from single two-dimensional layers 
parallel to the surface. Accordingly, the site index may be 
decomposed as $i=(i_\perp, i_\|)$ where $i_\perp$ labels the 
layers $i_\perp=1,\dots ,\infty$ and $i_\|$ the sites within a 
given layer. $i_\perp=1$ stands for the topmost surface layer. 
Within a tight-binding approach the hopping integrals $T_{ij}$ 
are assumed to be non-zero up to nearest neighbors only and 
constant throughout the system, i.\ e.: $T_{ij}=-t$ for nearest 
neighbors $i$, $j$. The energy scale is fixed by taking $t=1$. 
The energy zero will be chosen such that $T_{ii} \equiv T_0=0$. 
$U$ denotes the on-site Coulomb interaction which is assumed 
to be layer independent. A change of $U$ as well as of $T_{ij}$ 
in the vicinity of the surface should be expected for any real 
system \cite{utsurf}. Here the constancy of the parameters is 
a model assumption which helps to focus on the most important 
geometrical aspect, namely the reduction of the coordination 
number at the surface. Finally, in Eq.\ (\ref{eq:hubbard}) 
$n_{i\sigma}=c^\dagger_{i\sigma} c_{i\sigma}$ is the local 
occupation-number operator, and $\mu$ denotes the chemical 
potential.

All properties of the system that we are interested in can be 
obtained from the one-electron Green function:

\begin{equation}
  G_{ij\sigma}(E) = \langle \langle c_{i\sigma} ; 
  c^\dagger_{j\sigma} \rangle \rangle_E \: .
\label{eq:green}
\end{equation}
Let us assume the one-particle basis orbitals to be real. We 
then get the usual diagonal $i=j$ as well as the off-diagonal
$i\ne j$ quasi-particle density of states (QDOS) via:

\begin{equation}
  \rho_{ij\sigma}(E) = -\frac{1}{\pi} \mbox{Im} \, 
  G_{ij\sigma}(E+i0^+) \: .
\label{eq:qdos}
\end{equation}
For the semi-infinite system the QDOS will be layer dependent
in the vicinity of the surface. The Green function obeys an
equation of motion which in real-space representation reads:

\begin{equation}
  E G_{ij\sigma}(E) = \hbar \delta_{ij} + \sum_k \left(
  T_{ik} - \mu \delta_{ik} + \Sigma_{ik\sigma}(E) \right)
  G_{kj\sigma}(E) \: .
\label{eq:eom}
\end{equation}
Here we have introduced the electronic self-energy $\Sigma$ which 
includes all effects of electron correlations.

Within the usual perturbative approach \cite{pert} the self-energy
is expanded in powers of the interaction $U$. The first non-trivial
term is proportional to $U^2$. Using the real-space notation again,
we have:

\begin{equation}
  \Sigma_{ij\sigma}(E) = U \langle n_{i-\sigma} \rangle \delta_{ij}
  + U^2 \, \Sigma^{(2)}_{ij\sigma}(E) \: .
\label{eq:sigmaipt}
\end{equation}
The term linear in $U$ is the Hartree-Fock self-energy. 
Restricting ourselves to zero temperature, the second-order 
contribution reads:

\begin{eqnarray}
  && \hspace{-9mm} \Sigma^{(2)}_{ij\sigma}(E) = 
  \frac{1}{\hbar^3} 
  \! \int \!\!\! \int \!\!\! \int 
  \frac{\rho^{(0)}_{ij\sigma}(x) \rho^{(0)}_{ji-\sigma}(y) 
  \rho^{(0)}_{ij-\sigma}(z)}{E-x+y-z} \times
  \nonumber \\ &&
  (\Theta(-x) \Theta(y) \Theta(-z) + \Theta(x) \Theta(-y) \Theta(z))
  \: dx \, dy \, dz \: .
  \nonumber \\ &&
\label{eq:sigma2}
\end{eqnarray}
$\Theta$ denotes the step function. Within plain SOPT
$\rho^{(0)}$ is the free (off-diagonal) density of states:

\begin{equation}
  \rho^{(0)}_{ij\sigma}(E) = - \frac{1}{\pi} \mbox{Im} \, 
  G_{ij\sigma}(E+i0^+) \Big|_{U=0} \: .
\label{eq:rho1}
\end{equation}
The SOPT-HF approach (cf.\ e.~g.\ Refs.\ \cite{SC91,BJ90})
replaces the free by the Hartree-Fock density of states 
$\rho^{(1)}$ in the expression (\ref{eq:sigma2}):

\begin{equation}
  \rho^{(1)}_{ij\sigma}(E) = -\frac{1}{\pi} \mbox{Im} \, 
  G^{(1)}_{ij\sigma}(E+i0^+) \: ,
\end{equation}
where the HF Green function $G^{(1)}$ has to be calculated from

\begin{eqnarray}
  && \mbox{} \hspace{-7mm}  
  E \: G^{(1)}_{ij\sigma}(E) = \hbar \delta_{ij} 
  \nonumber \\
  && \mbox{} \hspace{0mm}  
  + \sum_k \Big(
  T_{ik} + (U \langle n_{i-\sigma} \rangle - \mu) \:
  \delta_{ik} \Big) \:
  G^{(1)}_{kj\sigma}(E) \: .
\label{eq:eomhf}
\end{eqnarray}
The ground-state expectation value of the local particle number
is obtained by integrating the (fully interacting) QDOS:

\begin{equation}
  \langle n_{i\sigma} \rangle =
  \frac{1}{\hbar}
  \int_{-\infty}^0 \rho_{ii\sigma}(E) \, dE \: .
\label{eq:nhf}
\end{equation}

Eqs.\ (\ref{eq:qdos}) -- (\ref{eq:nhf}) constitute a closed set 
of equations for the second-order perturbation theory around the 
Hartree-Fock solution. Generally, the local particle numbers
$\langle n_{i\sigma} \rangle$ have to be determined 
self-consistently. Due to the presence of the surface, 
the self-consistent values are expected to be layer dependent, 
i.~e.\ dependent on the distance from the surface. A self-consistent 
solution is not necessary in two cases, namely for an 
infinitesimally small interaction $U=0^+$ and for the 
symmetric case of a paramagnet at half-filling. In the latter, 
we have  
$\langle n_{i\sigma} \rangle = \langle n_{i-\sigma} \rangle= 0.5$ 
for all particle numbers, 
and the chemical potential is given by $\mu=U/2$.
(In the following the spin index will be omitted as far as possible; 
throughout the paper only paramagnetism is considered).

There are two main problems associated with the numerical 
solution of the SOPT-HF equations. The first one consists in 
the non-locality of the second-order term of the self-energy
(\ref{eq:sigma2}). For the calculation of the local contribution 
$i=j$ a three-dimensional energy integral has to be performed. 
Considering the imaginary part of the self-energy
$\mbox{Im} \, \Sigma^{(2)}_{ii}(E+i0^+)$, this 
reduces to convolution integrals which can be calculated
efficiently by applying a Laplace transform. The real part
is then obtained through the Kramers-Kronig relation. Details 
of this procedure can be found e.~g.\ in Refs.\ \cite{SC91,MH89b}.
Neglecting all off-diagonal elements in (\ref{eq:sigma2}) 
corresponds to the local approximation for the self-energy.
The numerical effort needed to account for the complete
non-locality of the self-energy can be estimated in the following
way: Let us consider the neighbor shells around the site $i$. 
If there was perfect translational invariance with respect to all 
three spatial dimensions, symmetry reasons would
require all non-local 
terms $\Sigma_{ij}$ to be equal for sites $j$ belonging to the same 
shell around the site $i$. Therefore, compared with the local 
approximation the effort rises by a factor $s+1$ that is 
determined by the number of shells $s$ that are necessary to 
obtain convergence for the results \cite{SC91} (as mentioned 
in the introduction, the necessary number of shells strongly 
depends on the lattice dimension $D$). 
In the semi-infinite system the symmetry is reduced, i.~e.\ 
not all the sites $j$ within the same shell are equivalent. 
Consider, for example, a (fixed) site $i$ near the surface. 
Independent from the lattice type and the surface there are 
three inequivalent possibilities at least 
to choose a site $j$ within the
nearest-neighbor shell of $i$: $j$ within the same layer as $i$
($j_\perp=i_\perp$), $j$ within a layer more distant from the 
surface ($j_\perp = i_\perp + n$ with $n>0$), and finally $j$ 
closer to the surface ($j_\perp = i_\perp - n$). (Note that
nearest neighbors need not necessarily belong to adjacent layers.) 
Therefore,
the reduced symmetry implies an increase in the numerical effort
by a factor three or even more, since the number
of inequivalent sites $j$ within the same shell around $i$ 
increases with increasing shell radius.
Finally, we have to consider all inequivalent positions for the
site $i$. Assuming perfect translational symmetry within the 
layers parallel to the surface, their number is given by the 
number of inequivalent layers $N_\perp$. The computational effort
therefore rises once more by a factor $N_\perp$. 
With increasing distance
of $i$ and $j$ to the surface, $\Sigma_{ij}(E)$ converges 
to the bulk self-energy. Thus $N_\perp$ can be regarded to be 
finite but may be 
rather large (several tens of layers). This also implies
that instead of the semi-infinite we can assume a slab geometry 
with a large number of layers $d$ ($\sim 2 N_\perp$) in the actual 
calculation.

The second main problem 
concerns the equation of motion (\ref{eq:eom})
for the Green function. It is the reduced translational symmetry
of the semi-infinite system again that renders its solution more 
difficult. In order to exploit translational symmetry within the
layers, two-dimensional Fourier transformation of any quantity 
$F_{ij}$ will be convenient:

\begin{equation}
  F_{ij} \equiv F_{i_\|j_\|}^{i_\perp j_\perp} =
  \frac{1}{N_\|} \sum_{{\bf k}_\|} 
  e^{i {\bf k}_\| ({\bf R}_{i_\|} - {\bf R}_{j_\|}) }
  F_{i_\perp j_\perp}({{\bf k}_\|}) \: ,
\label{eq:ft1}
\end{equation}
and
\begin{equation}
  F_{i_\perp j_\perp}({{\bf k}_\|}) =
  \frac{1}{N_\|} \sum_{i_\| j_\|} 
  e^{ - i {\bf k}_\| ({\bf R}_{i_\|} - {\bf R}_{j_\|}) }
  F_{i_\|j_\|}^{i_\perp j_\perp} \: .
\label{eq:ft2}
\end{equation}
Here ${\bf k}_\|$ is a wave vector of the first two-dimensional
surface Brillouin zone. After Fourier transformation the equation 
of motion for the Green function reads:

\begin{eqnarray}
  && \hspace{-10mm} 
  E \: G_{i_\perp j_\perp}({\bf k}_\|, E) = \hbar
  \delta_{i_\perp j_\perp} 
  + \sum_{k_\perp} 
  \Big( T_{i_\perp k_\perp}({\bf k}_\|) 
  \nonumber \\
  && \hspace{-2mm} 
  - \mu \delta_{i_\perp k_\perp} + 
  \Sigma_{i_\perp k_\perp}({\bf k}_\|, E) \Big) \:
  G_{k_\perp j_\perp}({\bf k}_\|, E) \: .
\label{eq:eomk}
\end{eqnarray}
This may be considered as a matrix equation for 
$G_{i_\perp j_\perp}$. The dimension of the inversion problem
is given by the slab thickness $d \sim 2 N_\perp$. 
Within the tight-binding 
approach, the hopping matrix $T_{i_\perp j_\perp}$ as well as
the self-energy matrix $\Sigma_{i_\perp j_\perp}$ are sparse. 
Therefore, the tight-binding recursion technique 
\cite{Hay80,PW84,PN96} is most
suitable for its solution. The application of the method yields the
Green function as a continued fraction:

\begin{eqnarray}
  && G_{i_\perp i_\perp}({\bf k}_\|, E) \equiv 
  G_{i_\perp i_\perp} =
  G^{(0)}_{i_\perp} \: ,
  \nonumber \\
  && G^{(k)}_{i_\perp} =
  \frac{\hbar}{E - a^{(k)}_{i_\perp}
  - \left(b^{(k+1)}_{i_\perp} \right)^2
  G^{(k+1)}_{i_\perp}} \: .
\label{eq:recursion}
\end{eqnarray}
The (energy-dependent)
coefficients $a^{(k)}_{i_\perp}$ and $b^{(k+1)}_{i_\perp}$ can
be calculated iteratively within the recursion approach. Due to the 
finite dimensionality, the continued fraction terminates 
automatically: $k=0,\dots d-1$. Off-diagonal elements of the 
Green function can be obtained by means of a polarization
relation \cite{PN96}.

{\center \bf \noindent III. RESULTS \\ \mbox{} \\}

We have chosen the simple-cubic (sc) lattice as a model system for
our calculations. Three different low-index surfaces, sc(100), 
sc(110) and sc(111), have been considered. Let us start the 
discussion of the results with the (100) face. For the sc lattice
the nearest-neighbor coordination number for sites in the bulk 
is $z=6$. Speaking in terms of the (100) layers parallel to the 
surface, each site has $z^{(100)}_\|=4$ nearest neighbors within 
the same layer and $z^{(100)}_\perp=2$ sites in the adjacent 
layers. For the top-layer sites ($i_\perp=1$) the coordination 
number is thus reduced to $z^{(100)}_S=5$. 

This (small) lessening of the coordination has consequences for 
the free ($U=0$) density of states (DOS) $\rho^{(0)}_{ij}(E)$. 
The free DOS determines via Eq.\ \ref{eq:sigma2} the second-order 
contribution of the self-energy for $U=0^+$. For finite $U$ 
and at half-filling we have to consider the Hartree-Fock (HF) 
DOS $\rho^{(1)}_{ij}(E)$. Compared with the free DOS the 
HF-DOS is only shifted constantly in energy by the amount 
$U \langle n_{i-\sigma} \rangle = U/2$ which is compensated by 
the corresponding shift of the chemical potential. So in both 
cases $\rho^{(0)}_{ij}(E)$ is the primary quantity of interest.

\begin{figure}[t] 
\unitlength1mm
\begin{picture}(0,98)
\put(0,-7){\epsfxsize=80mm \epsfbox{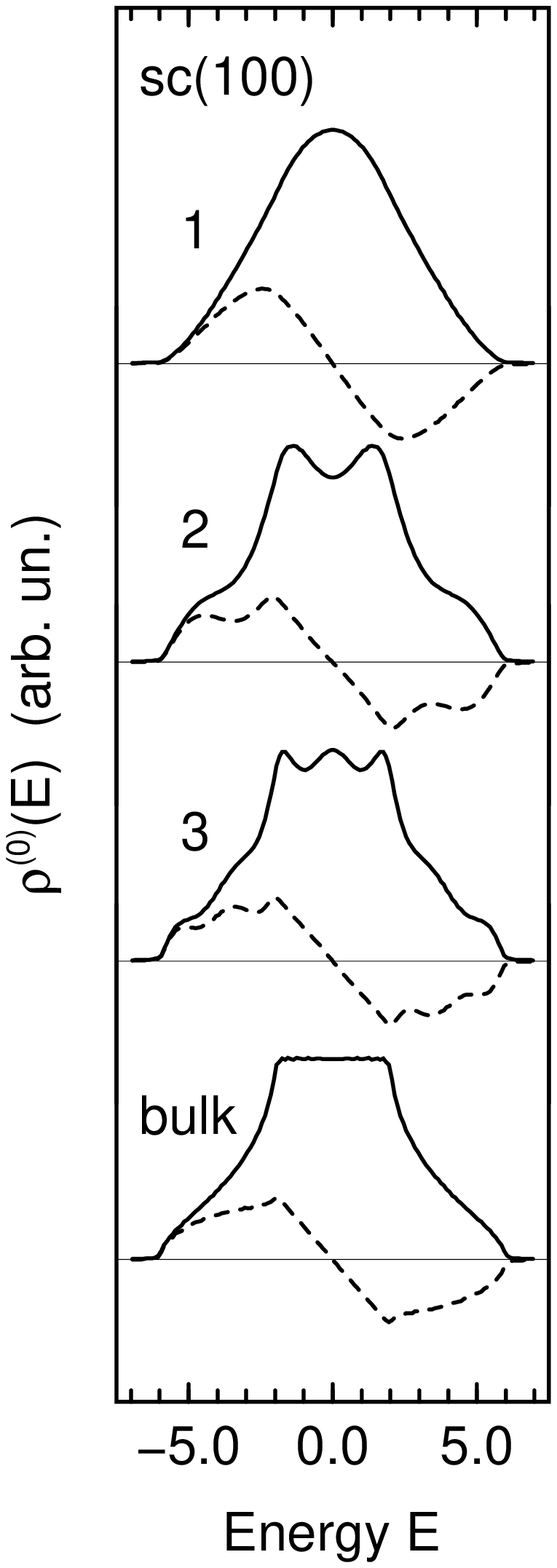}}
\end{picture}
\parbox[]{85mm}{\small Fig.~1.
Free ($U=0$) densities of states $\rho_{ii}^{(0)}$ for the 
first three layers from the surface ($i_\perp=1-3$) and free
bulk density of states ($i_\perp=\infty$) as functions of energy 
(solid lines). $i_\perp=1$ stands for the outermost surface layer.
Dashed lines: off-diagonal free densities of states
$\rho_{ij}^{(0)}(E)$ for nearest-neighbor sites $i$ and $j$, both
within the same (indicated) layer. Calculations for the sc(100)
surface. (A slab thickness of $d=50$ turns out to be sufficient for
convergence in all calculations).
All energies are given in terms
of the nearest-neighbor hopping $t=1$. $\mu=0$.
}
\end{figure}

Let us discuss first the layer-dependent diagonal elements ($i=j$) 
of the free DOS which are shown in Fig.~1. We notice that there 
are considerable differences between the DOS of the topmost surface 
layer and the bulk DOS. The difference becomes weaker and weaker 
with increasing distance from the surface. The surface can be 
regarded as a perturbation of the infinitely extended lattice 
that gives rise to oscillations of the layer-dependent charge 
density. These surface Friedel oscillations \cite{Mah90} are closely
related to the periodic deformations of the DOS which, as can be seen
in Fig.~1, are damped 
when passing from the very surface to the bulk. The layer-dependent 
off-diagonal DOS $\rho^{(0)}_{ij}(E)$ for nearest-neighbors $i$ 
and $j$ within the same layer is also shown in Fig.~1. Its
absolute magnitude is smaller, and there is a zero of the DOS
at the center of gravity $T_0=0$ of the diagonal DOS. Compared 
with the case $i=j$, we observe qualitatively similar differences 
between the top-layer DOS and the bulk DOS as well as the signature 
of surface Friedel oscillations.

The overall shape of the free DOS in the vicinity of the surface
can be understood in terms of the moments of the DOS. These
are defined as:

\begin{equation}
  M^{(n, 0)}_{ij} = \int_{-\infty}^\infty 
  E^n \rho_{ij}^{(0)}(E) \, dE \: .
\label{eq:mom1}
\end{equation}
An alternative, but equivalent representation can be derived from
the equation of motion for the free Green function:

\begin{equation}
  M^{(n, 0)}_{ij} = \sum_{i_1 \cdots i_{n-1}}
  T_{ii_1} T_{i_1i_2} \cdots T_{i_{n-1}j} \: .
\label{eq:mom2}
\end{equation}
Using this equation we can conclude that for each layer the 
diagonal ($i=j$) DOS must be a symmetric function of energy with
respect to its center of gravity $T_0=0$: All {\em odd} moments
must vanish since there is no way to start from and to return 
to a site $i$ by an odd number of nearest-neighbor hoppings 
for a (semi-infinite) simple cubic lattice. Analogously, the 
off-diagonal free DOS for nearest-neighbors $i$ and $j$ must 
be antisymmetric.

\begin{figure}[t] 
\unitlength1mm
\begin{picture}(0,90)
\put(-5,0){\epsfxsize=87mm \epsfbox{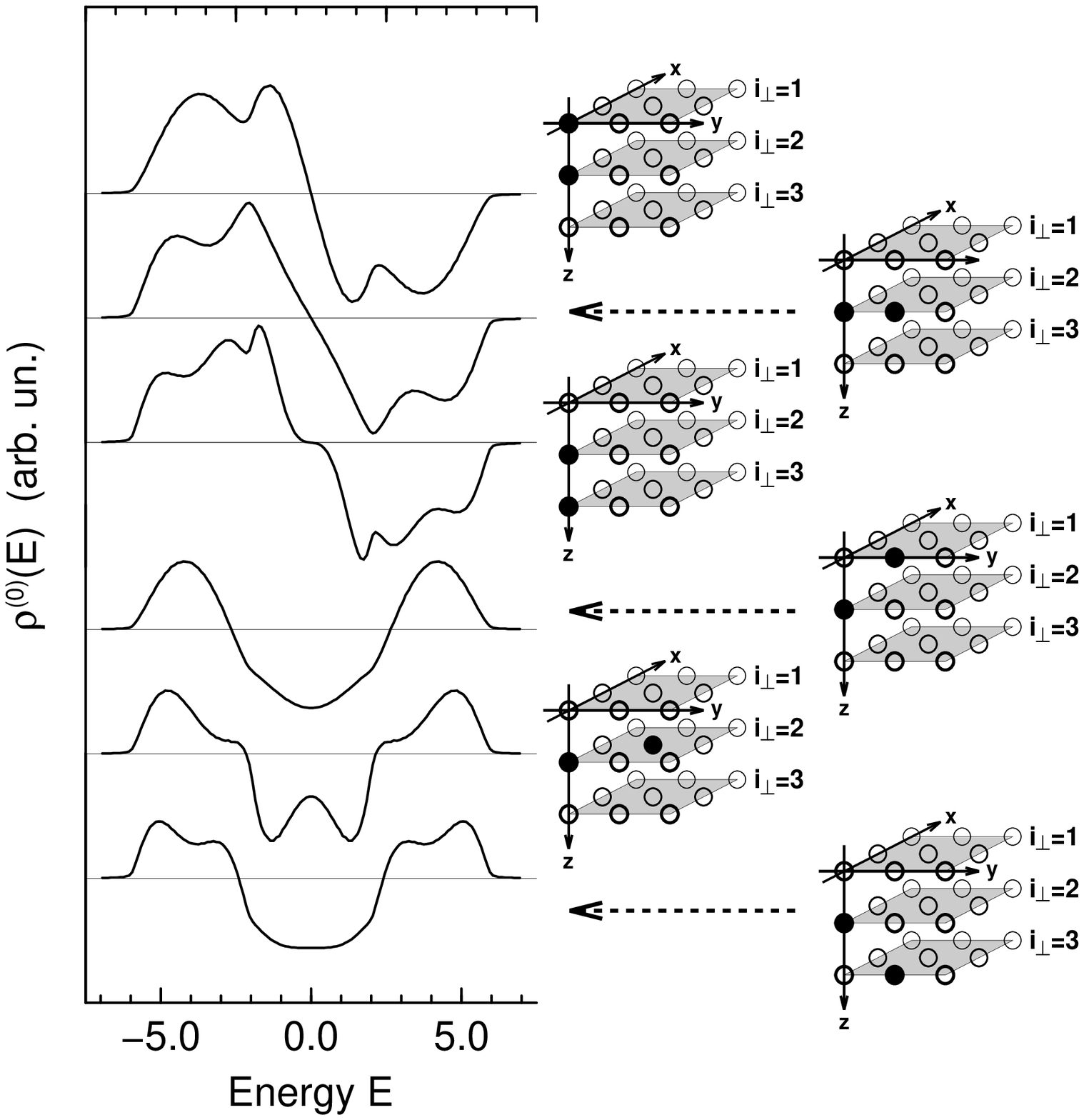}}
\end{picture}
\parbox[]{85mm}{\small Fig.~2.
Free off-diagonal densities of states $\rho_{ij}^{(0)}(E)$ near
the sc(100) surface. The respective sites $i$ and $j$ are
indicated as filled circles in the corresponding schematic drawing 
of the surface geometry. The shaded planes represent the first 
three layers from the surface, $i_\perp=1-3$ ($i_\perp=1$: top
surface layer). Note that $\rho_{ij}^{(0)}(E)=\rho_{ji}^{(0)}(E)$.
$\mu=0$.
}
\end{figure}

Comparing with the bulk, the on-site top-layer DOS is much 
more compressed. This ``band-narrowing'' is a consequence of the 
reduced coordination number of the top-layer sites and can be 
understood by considering the second moment:

\begin{equation}
  \Delta^2 \rho_{ii}^{(0)} = M_{ii}^{(2, 0)} - (M_{ii}^{(1, 0)})^2
  = \sum_{j\ne i} T_{ij}^2 = z(i) \: t^2 \: .
\end{equation}
The reduced coordination number $z(i)$ of the site $i$ at the
surface yields a reduced variance $\Delta^2 \rho^{(0)}$ of the 
top-layer
DOS. For the second and the third surface layer higher moments 
have to be considered. Despite the band-narrowing effect, the
densities of states of all layers share common band edges. For 
a spatially constant hopping parameter $t$ no split-off states 
can be observed.

The difference between the free densities of states for 
non-equivalent pairs of sites $i$ and $j$ is demonstrated
in Fig.~2. For the calculations the site $i$ has been kept
fixed within the second layer from the surface ($i_\perp=2$).
The first three panels from the top in Fig.~2 show the 
off-diagonal DOS $\rho^{(0)}_{ij}(E)$ for nearest-neighbor 
sites $j$. There are three non-equivalent positions relative to
$i$. We recognize significant though not strong differences 
between the spectra.
These are exclusively due to the presence of the surface; choosing
for $i$ a site within the bulk would yield identical results 
in all three cases. The last three panels show the DOS for $i$
within the second layer as before, but $j$ a next-nearest neighbor.
Again, three non-equivalent positions for $j$ can be found. 
Since an even number of nearest-neighbor hoppings is needed
to get from site $i$ to site $j$, the DOS is symmetric. The
differences between the spectra are non-negligible and of the 
same order of magnitude as in the nearest-neighbor case. They will
lead to a (weak) directional dependence of the self-energy in the 
vicinity of the surface.

\begin{figure}[t] 
\unitlength1mm
\begin{picture}(0,110)
\put(0,-7){\epsfxsize=80mm \epsfbox{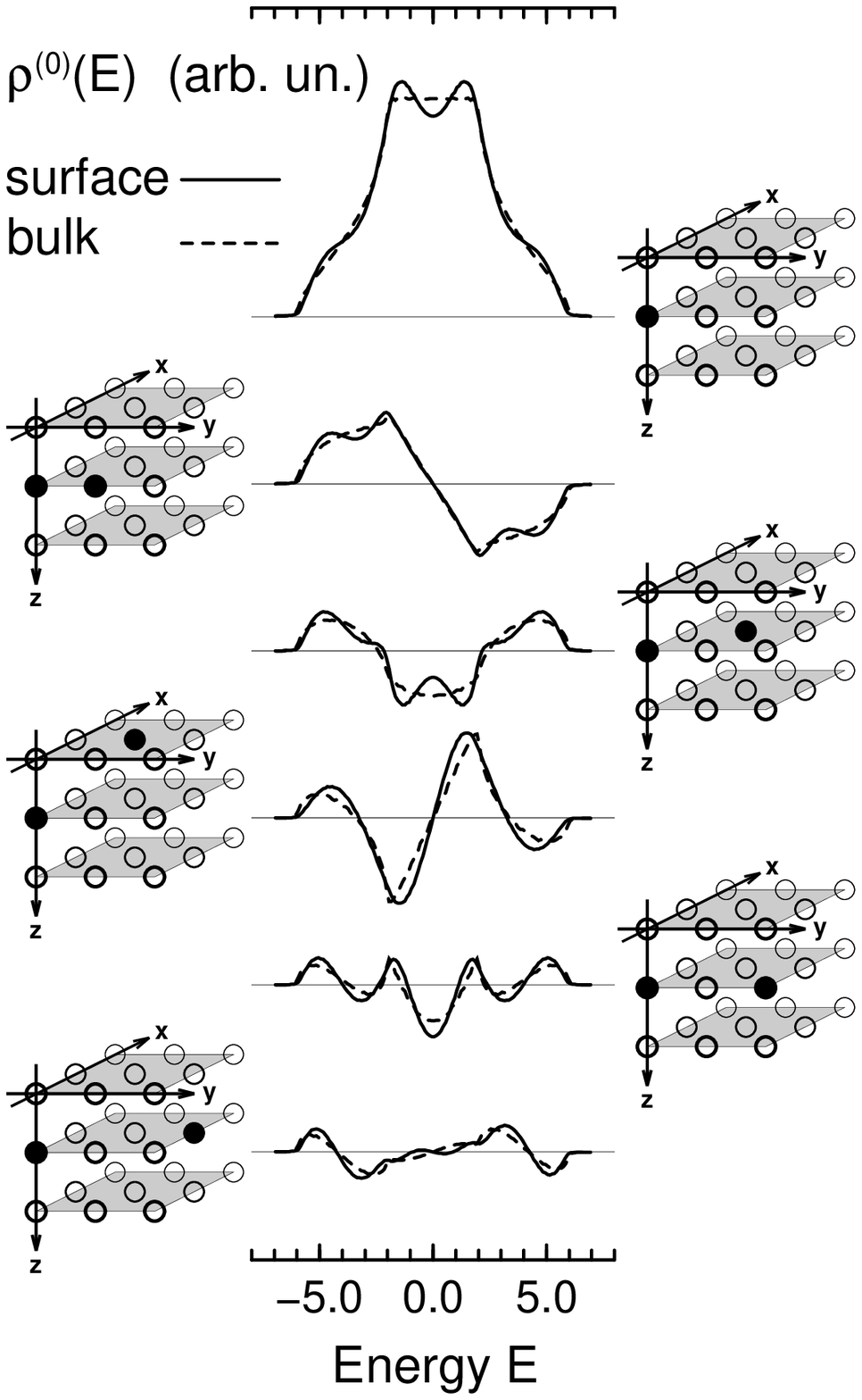}}
\end{picture}
\parbox[]{85mm}{\small Fig.~3.
Free densities of states $\rho_{ij}^{(0)}(E)$ near the sc(100)
surface (solid lines). Sites $i$ and $j$ are indicated (shaded
planes: (100) layers, top plane: surface layer). Dashed lines:
bulk densities of states. Relative position of sites $i$ and $j$ 
as indicated (shaded planes now to be interpreted as (100) layers
in the bulk). $\mu=0$.
}
\end{figure}

The local approximation for the self-energy $\Sigma_{ij}(E)$ 
can only be a reasonable starting point if the non-local parts 
strongly decrease with increasing distance between the sites 
$i$ and $j$. Within the perturbational approach these are 
determined by the off-diagonal free DOS. In Fig.~3 the DOS 
$\rho_{ij}^{(0)}(E)$ is shown for a site $i$ fixed within the 
second layer from the surface and different sites $j$ with an
increasing distance $d_{ij} \equiv ||{\bf R}_i - {\bf R}_j||$ 
(from the top). It can be seen that the absolute magnitude of 
the DOS decreases with increasing distance $d_{ij}$ indeed.
Particularly, the off-diagonal DOS for nearest neighbors is 
clearly smaller in absolute values compared with the 
on-site DOS over the whole energy range. A more quantitative
estimate can be given by integrating the absolute DOS. We then
obtain the ratio (for increasing distance, on-site DOS
normalized to 100) $100:33:21:28:14:9$. The decrease is not
monotonous. According to Eq.\ (\ref{eq:sigma2}) a stronger 
decrease has to be expected for the $U^2$-contribution to the 
self-energy since $\Sigma_{ij}^{(2)}$ (roughly)
results from the third power of the DOS $(\rho_{ij}^{(0)})^3$.

\begin{figure}[t] 
\unitlength1mm
\begin{picture}(0,60)
\put(6,-23){\epsfxsize=65mm \epsfbox{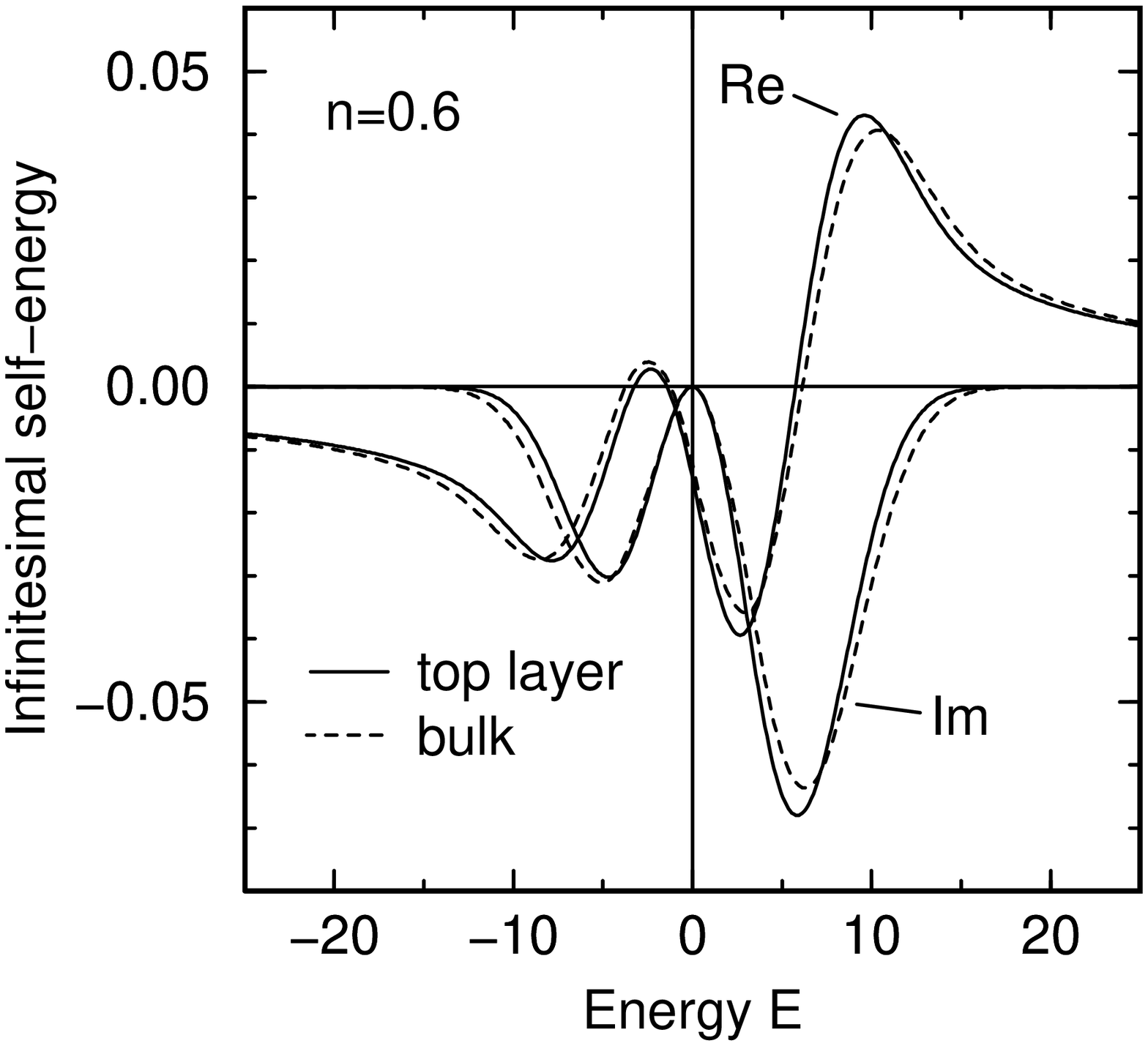}}
\end{picture}
\parbox[]{85mm}{\small Fig.~4.
Real and imaginary part of the infinitesimal on-site self-energy 
$\frac{1}{2} \frac{d^2}{dU^2} \Sigma_{ii}(E+i0^+)|_{U=0}$ of the 
semi-infinite sc(100) Hubbard model. Band-filling: $n=0.6$.
Solid line: site $i$ within the topmost surface layer 
($i_\perp=1$). Dashed line: bulk self-energy.
}
\end{figure}

The results for the site $i$ located near the surface can be 
compared with the results for a site $i$ in the bulk which are 
also shown in Fig.~3. The differences between the corresponding 
surface and bulk off-diagonal densities of states are significant 
though not strong. They are qualitatively similar to the 
difference between the surface and bulk on-site DOS. Since the 
second-order contribution 
to the self-energy (\ref{eq:sigma2}) results 
from a (two-fold) convolution of the free DOS, these differences 
become more or less meaningless. 
We thus can conclude that the self-energy
$\Sigma_{ij}^{(2)}(E)$ for a site $i$ within second layer and for an
arbitrary site $j$ is almost identical to the corresponding
diagonal or off-diagonal self-energy in the bulk. Therefore, 
for a site $i$
in the second surface layer the local approximation is likewise
appropriate (or inappropriate) as
for a site $i$ in the bulk.

Let us consider now the top layer where the situation is found 
to be different. Fig.~4 shows numerical results for the 
infinitesimal on-site self-energy 
$\frac{1}{2} \frac{d^2}{dU^2} \Sigma_{ii}(E)|_{U=0}$
of the semi-infinite sc(100) Hubbard model.
We consider a non-symmetric case: The chemical potential has 
been chosen such that in the bulk the average occupation number 
(band-filling) is $n=2\langle n_{i\sigma} \rangle = 0.6$. It can 
be seen in Fig.~4 that there are significant differences between 
the top-layer and the bulk self-energy. Contrary, the infinitesimal 
on-site self-energy of the second layer is almost indistinguishable 
from the bulk self-energy. On the energy scale used in the figure, 
differences would not be visible. The same holds true for the third 
and all other surface layers.

Fig.~4 shows that the on-site self-energy of the top layer is 
more compressed compared with the bulk self-energy; the effective 
width of the imaginary part is smaller at the surface. This is a
typical surface effect since it is a direct consequence of the 
narrowing of the top-layer DOS mentioned above. 
Another consequence concerns the local occupation number 
$\langle n_{i\sigma} \rangle$ at the surface. Due to the 
narrowing of the top-layer DOS, the surface charge density 
$n_{\rm s} \equiv 2 \langle n_{i\sigma} \rangle$ (for $i$ within 
the first layer) must be smaller than the bulk band-filling:
$n_{\rm s}<n$ (below half-filling) \cite{PN95}. For $n=0.60$ we find
$n_{\rm s}=0.56$. This leads to a reduced total weight
$w_{\rm s}$ of the imaginary part of the infinitesimal self-energy 
at the surface compared with its weight in the bulk $w$: 
From Eq.~(\ref{eq:sigma2}) we have:

\begin{equation}
w_{\rm (s)} \equiv 
- \frac{1}{\pi} \mbox{Im} \int_{-\infty}^\infty
\Sigma_{ii}^{(2)}(E+i0^+) \: dE =
\frac{n_{\rm (s)}}{2} \left(1-\frac{n_{\rm (s)}}{2}\right) \: .
\label{eq:w}
\end{equation}
The difference is found to be small: $w=0.21$ and $w_{\rm s}=0.20$.

Both, the imaginary part of the top-layer and of the bulk 
self-energy, vanish quadratically,
$-\mbox{Im} \Sigma_{ii}^{(2)}(E) \sim E^2$, as $E\mapsto 0$. 
This follows from a straightforward analysis of Eq.~(\ref{eq:sigma2})
\cite{Lut61} and implies the existence of a well-defined Fermi 
surface \cite{Lut60}. We can thus distinguish between the damping 
of the occupied ($E<0$) and of the unoccupied ($E>0$) part in the 
excitation spectrum. Let 
$w_{\rm (s)}^<$ and $w_{\rm (s)}^>$ denote the integrated weight 
of $-\frac{1}{\pi} \mbox{Im} \Sigma_{ii}^{(2)}(E+i0^+)$ at the 
surface or in the bulk for $E<0$ and $E>0$, respectively. We have 
$w_{\rm (s)}^<+w_{\rm (s)}^>=w_{\rm (s)}$, and from 
Eq.~(\ref{eq:sigma2}) the following result can be derived:

\begin{eqnarray}
   w_{\rm (s)}^< &=& 
   \left( \frac{n_{\rm (s)}}{2} \right)^2 
   \left(1-\frac{n_{\rm (s)}}{2}\right) \: ,
\nonumber \\
   w_{\rm (s)}^> &=& 
   \frac{n_{\rm (s)}}{2} \left(1-\frac{n_{\rm (s)}}{2}\right)^2 \: .
\label{eq:ouw}
\end{eqnarray}
It can be seen in Fig.~4 that for the occupied part the integrated
weight is clearly smaller in the top layer. This has to be 
interpreted as another typical surface effect: The reduction 
of the coordination number at the surface implies the 
narrowing of the top-layer DOS and thus the lowered charge 
density; via Eq.~(\ref{eq:ouw}) this leads to a smaller damping 
in the occupied part. We get $w_{\rm s}^<=0.056$ to be 
compared with the bulk value $w^<=0.063$. On the contrary, 
the weight within the unoccupied part is almost unchanged:
$w_{\rm s}^>=0.145$ and $w^>=0.147$. 

The non-local parts of the infinitesimal self-energy for the top
layer of the sc(100) surface and for the bulk are shown in Fig.~5.
Unlike the imaginary part of on-site (retarded) self-energy,
which must be negative over the whole energy range, there is no 
definite sign for the imaginary part in the off-diagonal case.
Instead, we have the sum rule $\int^\infty_{-\infty} \mbox{Im} \, 
\Sigma_{ij}^{(2)}(E+i0^+) \, dE = 0$ for $i\ne j$. The low-energy 
behavior, however, remains unchanged:
the imaginary part vanishes quadratically for $E\mapsto 0$, i.~e.\
$\mbox{Im} \, \Sigma^{(2)}_{ij}(E) \sim \pm E^2$ for $i\ne j$. 
The sign depends
on the sign of the (off-diagonal) DOS at the Fermi edge. This can
be verified by expanding the infinitesimal self-energy in powers 
of $E$. Similar to the analysis in Ref.\ \cite{Lut61}, one may 
derive the following result in the real-space representation:

\begin{equation}
  -\frac{1}{\pi} \mbox{Im} \: \Sigma_{ij}^{(2)}(E+i0^+) =
  \left(\frac{1}{\hbar} 
  \rho_{ij}^{(0)}(0) \right)^3 \times E^2 \: + \: {\cal O}(E^4) \: .
\end{equation}
Consider, for example, nearest-neighbor sites $i$ and $j$ and
$n<1$. As can be seen from Fig.~1, 
$\rho_{ij}^{(0)}(E)$ is positive at the Fermi energy and thus 
$\mbox{Im} \, \Sigma_{ij}^{(2)}(E+i0^+) \sim - E^2$ for $E\mapsto 0$
which is consistent with the result shown in Fig.~5.

\begin{figure}[t] 
\unitlength1mm
\begin{picture}(0,115)
\put(0,-2){\epsfxsize=80mm \epsfbox{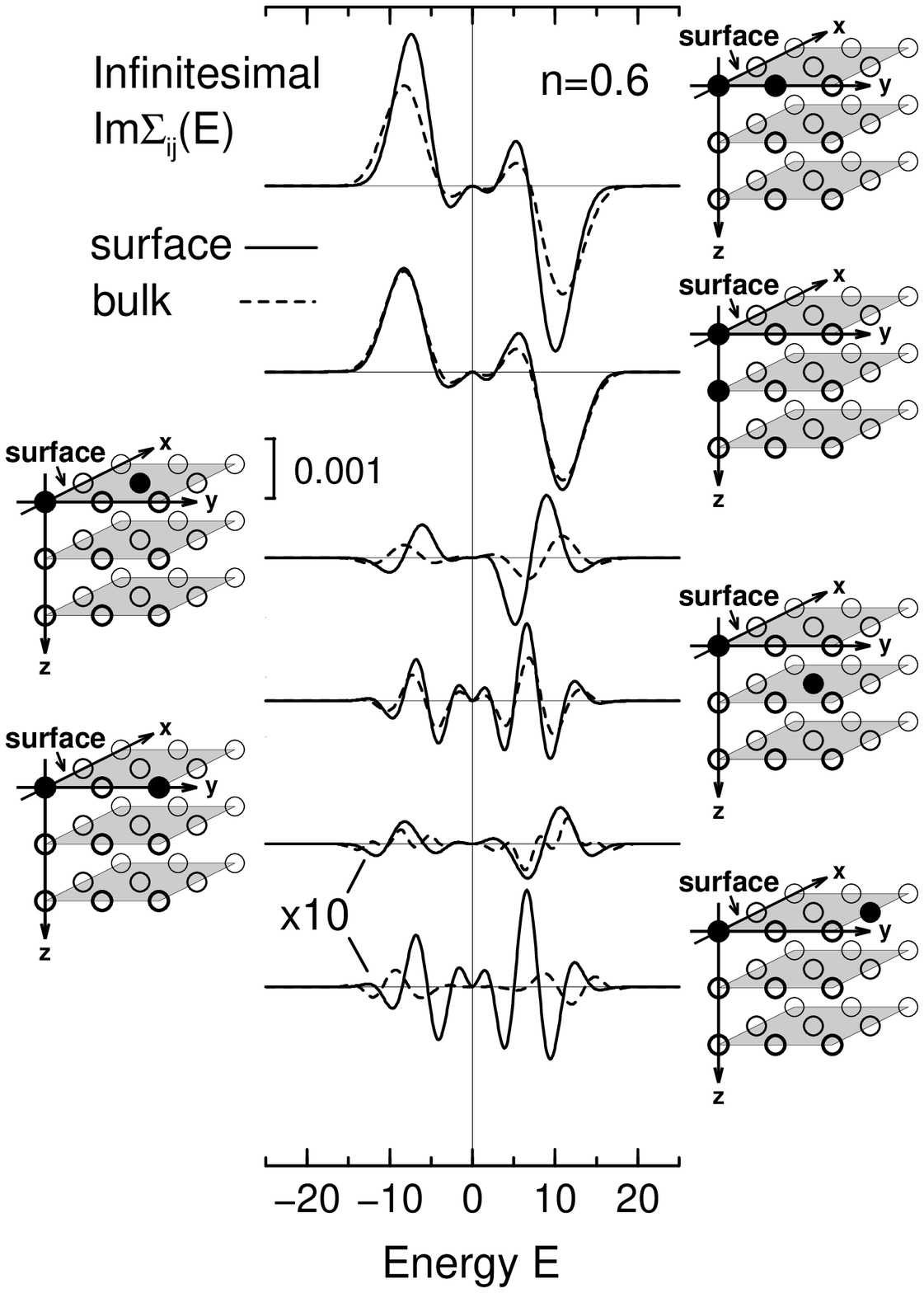}}
\end{picture}
\parbox[]{85mm}{\small Fig.~5.
Off-diagonal elements of the imaginary part of the infinitesimal
self-energy $\frac{1}{2} \frac{d^2}{dU^2} \Sigma_{ij}(E+i0^+)|_{U=0}$
near the sc(100) surface (solid lines) and in the bulk (dashed 
lines). Band-filling $n=0.6$. Sites $i\ne j$ as indicated (dashed 
lines/bulk: shaded planes represent (100) {\em bulk} layers). 
The vertical energy scale is given by the height of the bar 
($0.001$ in units of $t$).
In the last two panels the data have been multiplied by a factor 10.
}
\end{figure}

The first two panels from the top in Fig.~5 show the imaginary part
of the self-energy for two non-equivalent pairs of nearest-neighbor 
sites $i$ and $j$. A significantly stronger variation of the
self-energy is observed when both sites are located within the 
top layer. This implies a directional dependence of the self-energy
which, however, is confined to the very surface. In the bulk both
cases are equivalent, and the same result for the self-energy is
found. Generally we can state that
significant though not strong differences to the bulk self-energy 
are only found in those cases where both sites $i$ and $j$ belong 
to the topmost surface layer. This again proves that the presence 
of the surface manifests itself in the self-energy for the top 
surface layer only. 

\begin{figure}[t] 
\unitlength1mm
\begin{picture}(0,120)
\put(7,12){\epsfxsize=70mm \epsfbox{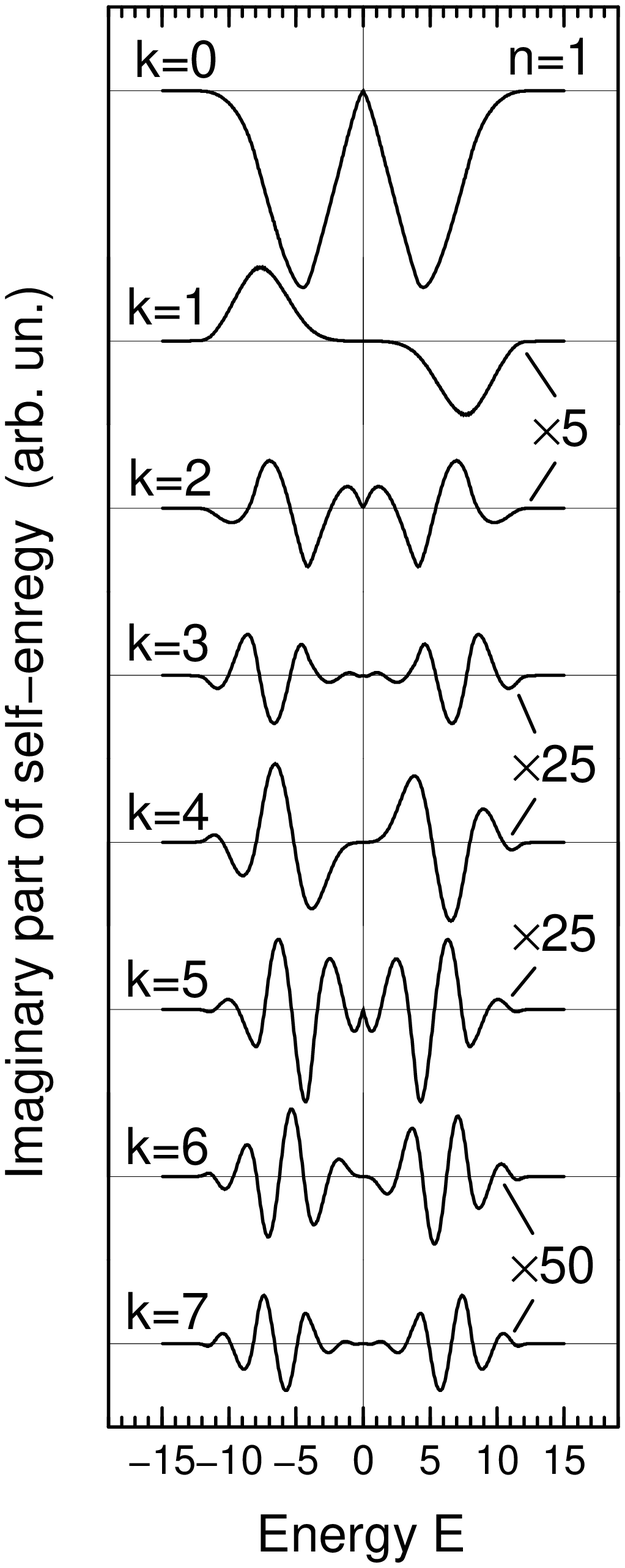}}
\end{picture}
\parbox[]{85mm}{\small Fig.~6.
Imaginary part of the (infinitesimal) self-energy 
$\frac{1}{2} \frac{d^2}{dU^2} \Sigma_{ij}(E+i0^+)|_{U=0}$ for the 
two-dimensional ($D=2$) Hubbard model on the square lattice
at half-filling ($n=1$). Site $j$ within the $k$-th neighbor
shell of the (fixed) site $i$. Results for $k=0$ (on-site
self-energy) up to $k=7$. The results for $k\ne 0$ have been scaled 
as indicated.
}
\end{figure}

Increasing the distance $d_{ij}$ between the sites, results in a
rapid decrease of the absolute values for the self-energy in the 
bulk as well as in the vicinity of the surface. Compared with the 
on-site self-energy (Fig.~4), the absolute values of the non-local
self-energy are smaller by more than one order of magnitude already
for nearest-neighbor sites $i$ and $j$. We have integrated the 
energy-dependent expression 
$|\mbox{Im} \, \Sigma_{ij}^{(2)}(E+i0^+)|$ 
for $i=j$ (Fig.~4) and for $i\ne j$ (second to sixth panel in Fig.~5)
to get a more quantitative estimate for the magnitude of the
different contributions at the very surface. Normalizing the 
local contribution to 100, the following ratio is obtained for 
increasing distance: $100:4.3:1.4:1.6:0.09:0.20$. As had to be
expected, this roughly corresponds to the above-mentioned ratio 
for the third powers of the integrated (absolute) densities of 
states. It shows up that the local part of the self-energy 
(bulk as well as surface) is rather dominating. 

The results for the semi-infinite lattice may be compared with
those for a two-dimensional case. For this purpose we have 
calculated the local and the non-local parts of the infinitesimal 
self-energy on the $D=2$ square lattice. Fig.~6 shows the results
for the imaginary part at half-filling. The self-energy 
$\Sigma_{ij}^{(2)}(E)$ is a symmetric (antisymmetric) function 
of energy with respect to $E=0$ if an even (odd) number of 
nearest-neighbor hoppings is necessary to get from site $i$ 
to site $j$. At half-filling the low-energy behavior of the 
self-energy is likewise exceptional: Expanding the expression
(\ref{eq:sigma2}) for small $E$ yields for the on-site part:
$-\frac{1}{\pi} \mbox{Im} \, \Sigma_{ii}^{(2)}(E+i0^+) \sim 
E^2 |\ln |E||^3$ as $E\mapsto 0$ \cite{edisc}. The reason
for this anomalous behavior at half-filling is the logarithmic 
divergence of the free DOS at $E=0$ in two dimensions. 

As can be seen from Fig.~6, the non-local parts of the self-energy 
are much smaller in absolute magnitude compared with the local one.
Again they are found to decrease with increasing distance $d_{ij}$.
For the integrated absolute self-energies we obtain the following
ratio: $100:10:7.4:1.1:2.1:2.0:0.80:0.52$ (on-site part normalized 
to 100). Compared with the results found for the sc(100) surface, 
we notice a considerably slower decrease with increasing distance 
$d_{ij}$. Qualitatively, this result does not depend on the 
band-filling at all. Although the mere numbers should not be 
overemphasized, the comparison of the different ratios surely 
gives an impression about the relative importance of the 
non-locality of the self-energy in the different cases. 

\begin{figure}[t] 
\unitlength1mm
\begin{picture}(0,115)
\put(3,12){\epsfxsize=70mm \epsfbox{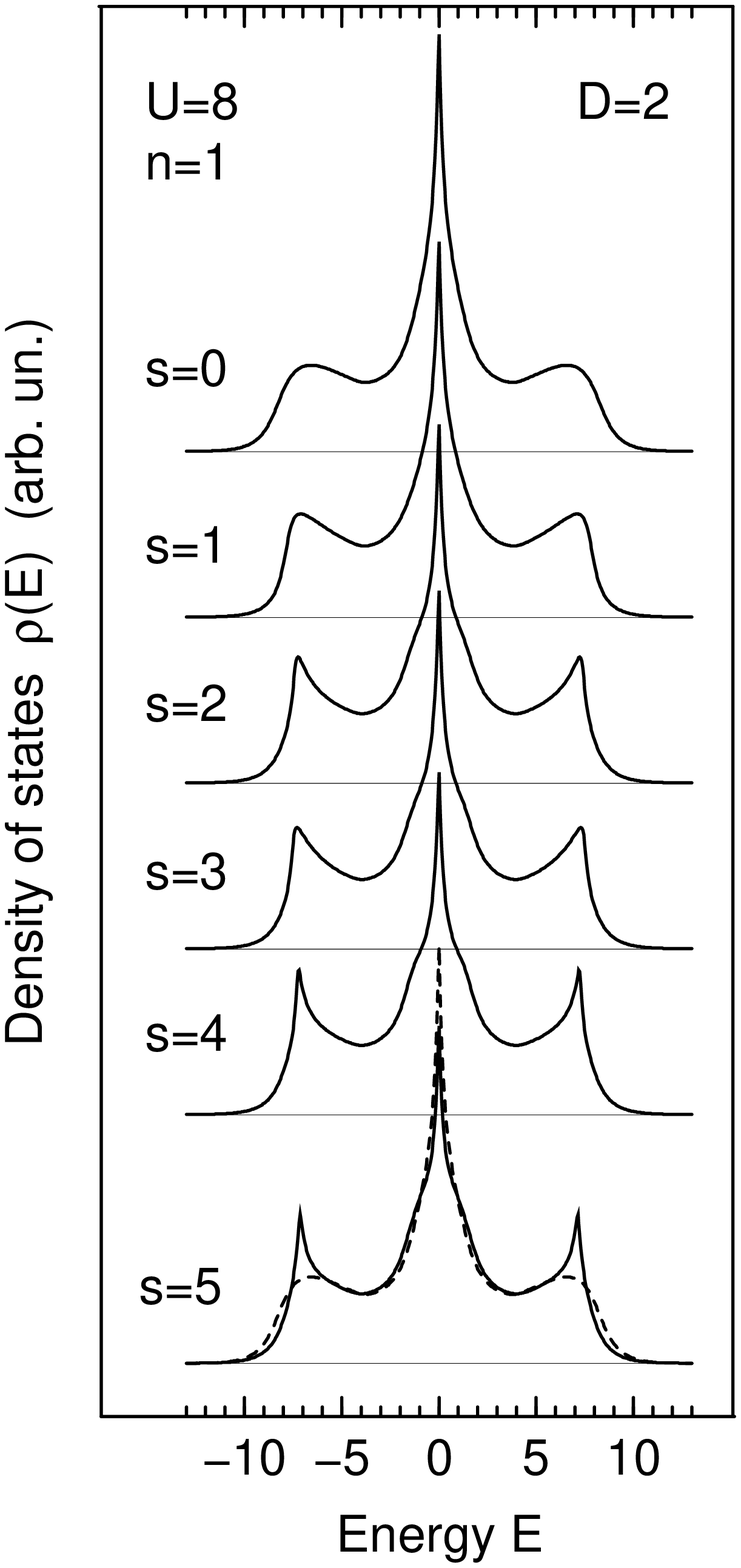}}
\end{picture}
\parbox[]{85mm}{\small Fig.~7.
Quasi-particle density of states $\rho(E)$ of the $D=2$ Hubbard 
model on the square lattice at half-filling and $U=8$ within SOPT-HF.
Off-diagonal contributions of the self-energy included up to the 
s-th neighbor shell. Results for $s=0$ (local approximation) up to 
$s=5$. Dashed line: $s=0$ result for comparison.
}
\end{figure}

So far only the free DOS as well as the self-energy have been
considered. To be able to judge on the quality of the local 
approximation, however, one also has to check the effects of the 
non-local parts on the quasi-particle density of states (QDOS):
While $|\Sigma_{ij}^{(2)}(E+i0^+)|$ may be a small quantity for a 
given pair of sites $i\ne j$, it nevertheless may be important with 
respect to the QDOS provided that a large number of equivalent 
pairs have to be taken into account. For the discussion of 
the (paramagnetic) QDOS at finite interaction $U$, we restrict 
ourselves to the symmetric case of half-filling. As mentioned 
in the introduction, the SOPT-HF scheme reproduces a large 
number of exactly solvable limits for $n=1$ and should thus 
yield reliable results, at least on a qualitative level. 

The QDOS of the $D=2$ square lattice at half-filling is shown
in Fig.~7. We have chosen $U=8$ which is equal to the free Bloch
band width $W$. The first panel shows the result for the local
approximation ($s=0$). The other panels show calculations where we 
have included the non-local parts of the self-energy up to $s$-th 
neighbor shell ($s=1$ to $s=5$). Qualitatively, all spectra look 
similar: The high-energy charge excitation peaks (Hubbard bands) 
are clearly visible for $U=8$, and a quasi-particle resonance 
remains
around $E=0$. Yet there are important differences: The inclusion
of more and more shells narrows the whole spectrum, sharpens the 
charge-excitation peaks while the resonance looses some spectral 
weight. Furthermore, a new structure comes into existence at 
$E\approx \pm 1$. Fully converged results cannot be obtained
until $s=5$ \cite{tpdisc}. 

\begin{figure}[t] 
\unitlength1mm
\begin{picture}(0,105)
\put(9,0){\epsfxsize=70mm \epsfbox{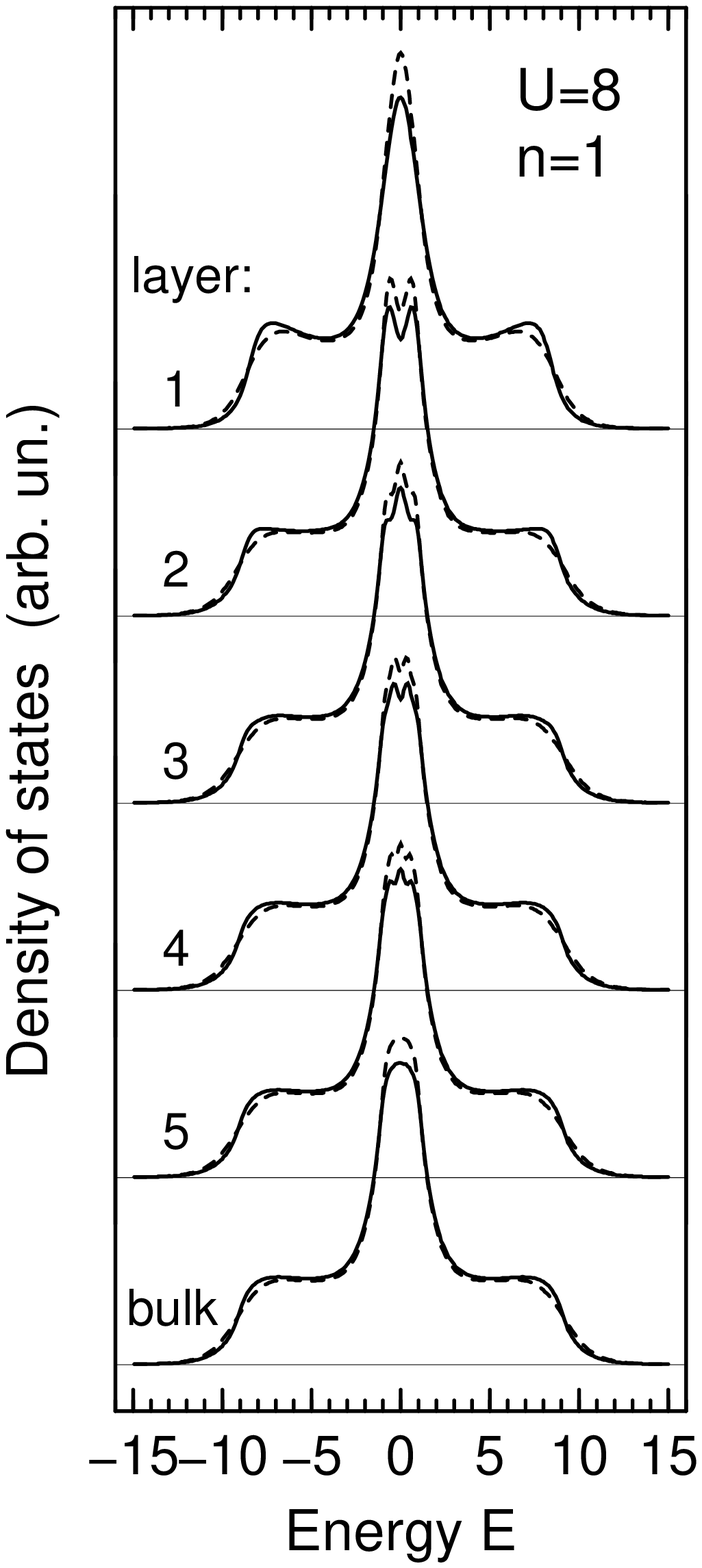}}
\end{picture}
\parbox[]{85mm}{\small Fig.~8.
Layer-dependent quasi-particle density of states $\rho_{ii}(E)$
for the semi-infinite sc(100) Hubbard model at half-filling and
$U=8$. SOPT-HF results for the first five layers from the surface
and for the bulk. Dashed lines: local approximation. Solid lines:
non-locality of the self-energy fully taken into account.
}
\end{figure}

Let us now return to the sc(100) surface again. The 
layer-dependent QDOS in the vicinity of the surface and in 
the bulk for $U=8$ and at half-filling are shown Fig.~8. For 
each layer the QDOS exhibits three main features which again 
can be interpreted as the two charge-excitation peaks (on the 
high-energetic sides) and a resonance (at the Fermi level). 

The discussion of the self-energy above has already indicated 
that effects of its non-locality are less important for the 
semi-infinite system compared with $D=2$. This becomes 
manifest when considering the QDOS at the surface. Taking into 
account the on-site self-energy and the nearest-neighbor non-local 
part is almost sufficient to obtain fully convergent results. 
Remaining differences between the QDOS for $s=1$ and for higher 
$s$ would hardly be visible on the scale used in Fig.~8. Let us 
point out that the convergence with respect to $s$ is equally fast 
for both, the top-layer and the bulk QDOS \cite{updisc}.

The results shown in the figure prove that the local approximation 
provides a reasonable description of all features appearing in the 
spectra. Only the spectral weight of the resonance at the Fermi 
edge is somewhat overestimated, and the charge-excitation peaks 
are less pronounced for $s=0$. Compared with the results for the 
$D=2$ square lattice, however, the local approximation is only 
slightly better. In any case it is likewise appropriate for the 
bulk as well as for the surface including the top layer.
 
\begin{figure}[t] 
\unitlength1mm
\begin{picture}(0,72)
\put(5,-24){\epsfxsize=70mm \epsfbox{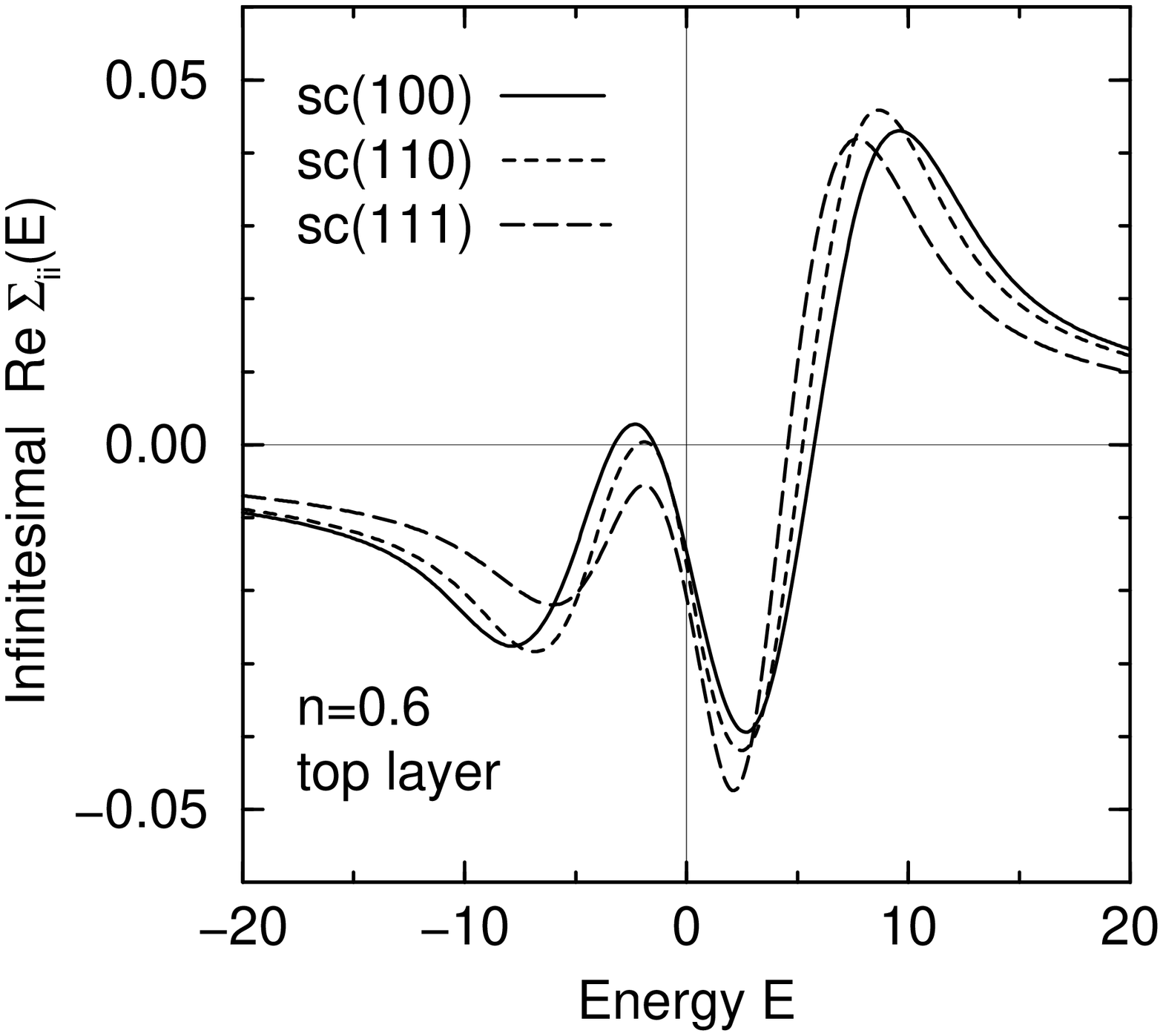}}
\end{picture}
\parbox[]{85mm}{\small Fig.~9.
Real part of the infinitesimal on-site self-energy 
$\frac{1}{2} \frac{d^2}{dU^2} \Sigma_{ii}(E+i0^+)|_{U=0}$
for the top surface layer of the semi-infinite Hubbard model.
Results for three different surfaces. Band-filling: $n=0.6$.
}
\end{figure}

\begin{figure}[t] 
\unitlength1mm
\begin{picture}(0,106)
\put(6,7){\epsfxsize=65mm \epsfbox{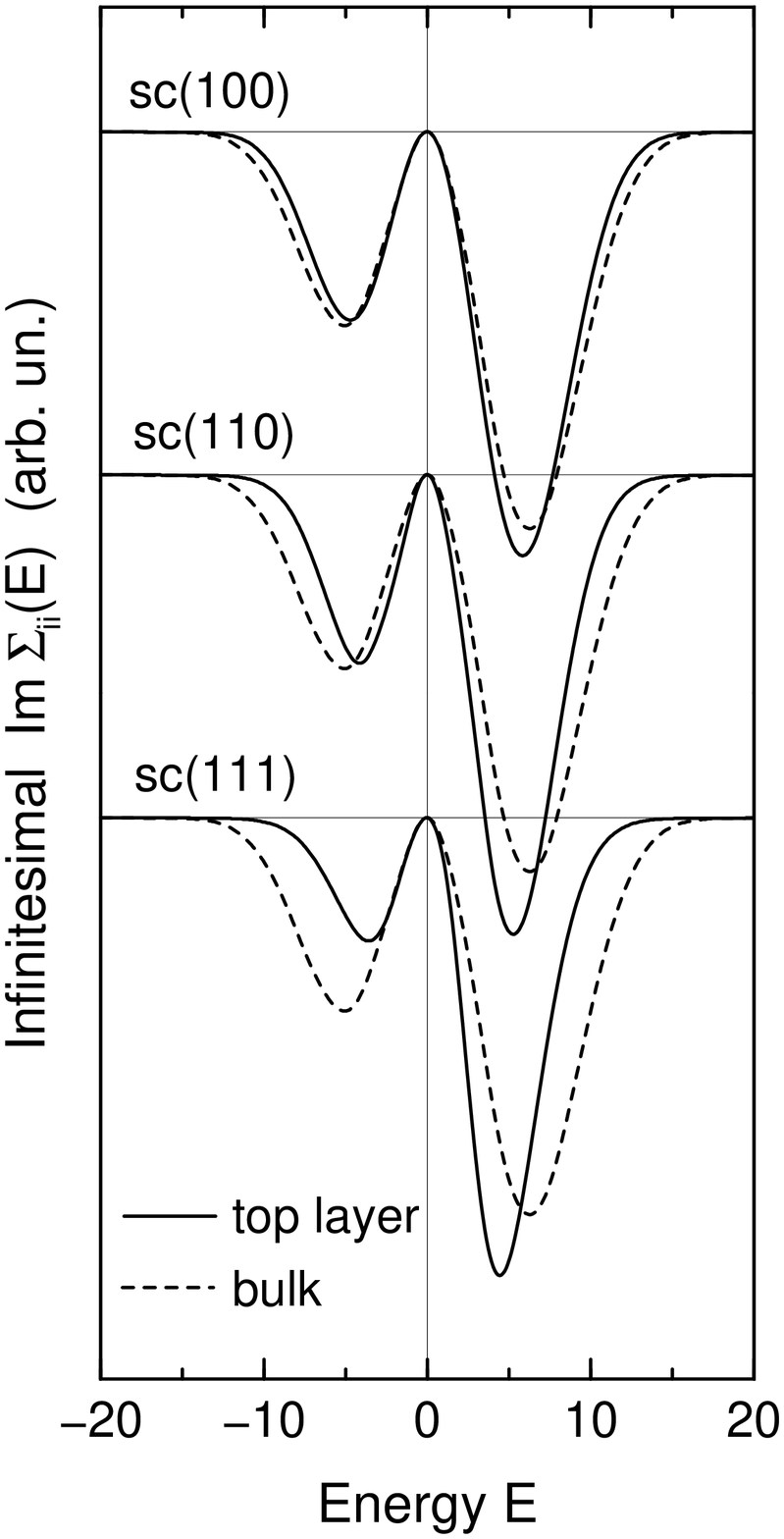}}
\end{picture}
\parbox[]{85mm}{\small Fig.~10.
Imaginary part of the infinitesimal on-site self-energy 
$\frac{1}{2} \frac{d^2}{dU^2} \Sigma_{ii}(E+i0^+)|_{U=0}$
of the semi-infinite Hubbard model at $n=0.6$. Solid lines:
site $i$ within the top surface layer. Dashed lines: bulk.
}
\end{figure}

The results for the QDOS at the sc(100) surface are interesting
of their own: First, we notice that the top-layer QDOS is slightly
narrowed. This can be understood by referring to the moments again. 
The first few moments of the QDOS,

\begin{equation}
  M^{(n)}_{ij} = \int_{-\infty}^\infty 
  E^n \rho_{ij}(E) \, dE \: ,
\label{eq:qmom1}
\end{equation}
can be calculated exactly (see Ref.\ \cite{PN96}, for example):

\begin{eqnarray}
  M^{(0)}_{ii} & = & 1
\nonumber \\
  M^{(1)}_{ii} & = & T_0 + U \langle n_{i-\sigma} \rangle
\nonumber \\
  M^{(2)}_{ii} & = & \sum_j T_{ij}^2 + 2 U T_0
  \langle n_{i-\sigma} \rangle + 
  U^2 \langle n_{i-\sigma} \rangle \: .
\label{eq:qmom2}
\end{eqnarray}
The variance of the QDOS for the layer $i_\perp$ is then given 
by the expression:

\begin{equation}
  \Delta^2 \rho_{ii} = M_{ii}^{(2)} - (M_{ii}^{(1)})^2
  = \sum_{j\ne i} T_{ij}^2 + U^2 
  \langle n_{i-\sigma} \rangle 
  (1 - \langle n_{i-\sigma} \rangle) \: .
\label{eq:var}
\end{equation}
We can conclude that increasing $U$ leads to a strong increase of
the variance of the QDOS which is mainly due to the evolution of
the two Hubbard bands. For a given $U$ and at half-filling the 
second term on the right-hand side of (\ref{eq:var}) is a constant.
As for the case of the free DOS, the first term implies a narrowing
of the top-layer QDOS because of the reduced coordination number
for sites in the first layer. This can be seen in Fig.~8. 
Additionally, we notice a slight
narrowing of the quasi-particle resonance for the top layer.

Compared with the free DOS in Fig.~1, the differences between the 
QDOS for the different layers are much smaller in general. This is
a consequence of quasi-particle damping due to the imaginary
part of the self-energy. Generally this indicates a damping of the 
surface Friedel oscillations due to electron correlations. Note 
that the effect is also reproduced by the local approximation.
The strong periodic deformations of the 
DOS in the vicinity of the surface that could be observed for the 
free ($U=0$) system (Fig.~1), are now confined to the resonance 
in small energy range around the Fermi energy. Here they appear as
tiny wiggles only.

\begin{figure}[t] 
\unitlength1mm
\begin{picture}(0,100)
\put(6,0){\epsfxsize=65mm \epsfbox{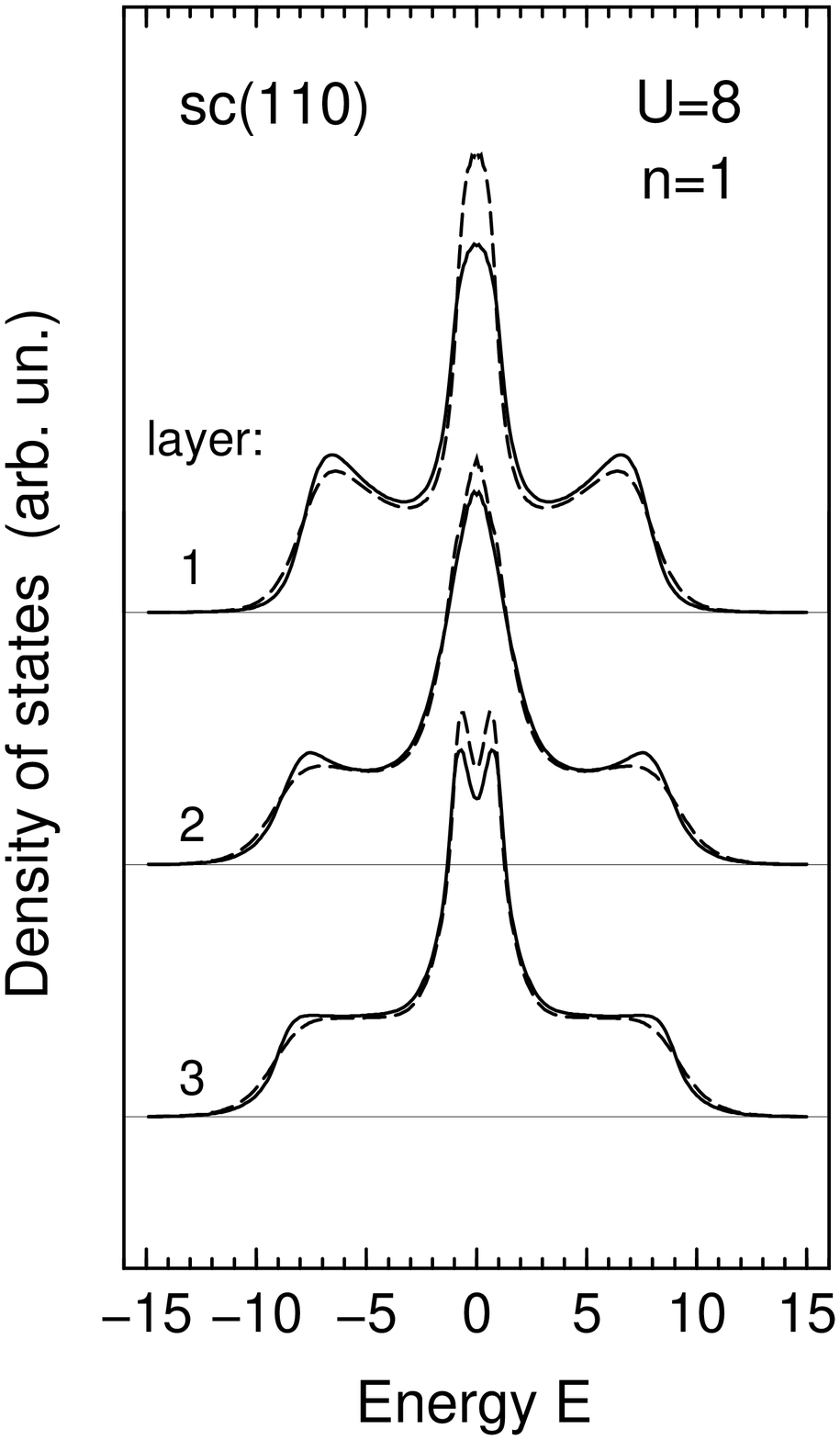}}
\end{picture}
\parbox[]{85mm}{\small Fig.~11.
Layer-dependent quasi-particle density of states for the sc(110)
surface at half-filling and $U=8$. SOPT-HF results for the first
three layers from the surface. Dashed lines: local approximation
for the self-energy. Solid lines: non-locality of self-energy 
fully taken into account.
}
\end{figure}

The effects that result from the reduction of the coordination
number at the surface show up more clearly when comparing results
for different surfaces. For this purpose we also considered 
the (110) and the (111) surface of the sc lattice. Thinking 
the respective system to be built up from layers parallel to the 
surface, each site except for those in the first layer has 
$z^{(110)}_\|=2$ ($z^{(111)}_\|=0$) nearest neighbors within the 
same and $z^{(110)}_\perp=4$ ($z^{(111)}_\perp=6$) nearest neighbors
in the adjacent layers. This implies $z^{(110)}_S=4$ and
$z^{(111)}_S=3$ for the coordination numbers of the top-layer 
sites, and we have $z^{(100)}_S > z^{(110)}_S > z^{(111)}_S$.

This sequence can be recovered in the results for the top-layer 
self-energy for the different surfaces. Fig.~9 shows their
real parts for $U=0^+$ and bulk band-filling $n=0.6$. The 
comparison shows significant differences which prove the 
top-layer self-energy to be sensitively dependent on the 
type of the surface. The more open the surface, the stronger 
is the narrowing of the top-layer free DOS and thus the narrowing
of the self-energy. In Fig.~9 this effect can be seen most
clearly at high energies. 

It is even more apparent in Fig.~10 where the corresponding 
imaginary parts are shown in comparison with the imaginary part 
of the bulk self-energy. The strongest differences with respect 
to the bulk are found for the (111) surface. As has been 
mentioned above, the lowered variance of the top-layer free
DOS implies a reduced charge density at the surface:
$n_{\rm s}<n<1$. We find $n_{\rm s}^{(111)}=0.38$ and 
$n_{\rm s}^{(110)}=0.52$ to be compared with
$n_{\rm s}^{(100)}=0.56$ and $n=0.6$. 
According to Eq.\ (\ref{eq:w}) this
results in a diminished total weight of the imaginary part of the
top-layer infinitesimal self-energy. Again the weight is reduced
in the occupied part $E<0$ mainly.

In the case of the (100) surface, differences between the local 
self-energy at the surface and in the bulk are worth mentioning 
for the top layer only (as discussed above). The same holds for 
the (110) surface. In the case of the (111) face there are 
non-negligible differences also for the second layer, while the
local self-energy of the third (and all other) surface layer is
almost indistinguishable from the bulk self-energy.

\begin{figure}[t] 
\unitlength1mm
\begin{picture}(0,100)
\put(6,8){\epsfxsize=65mm \epsfbox{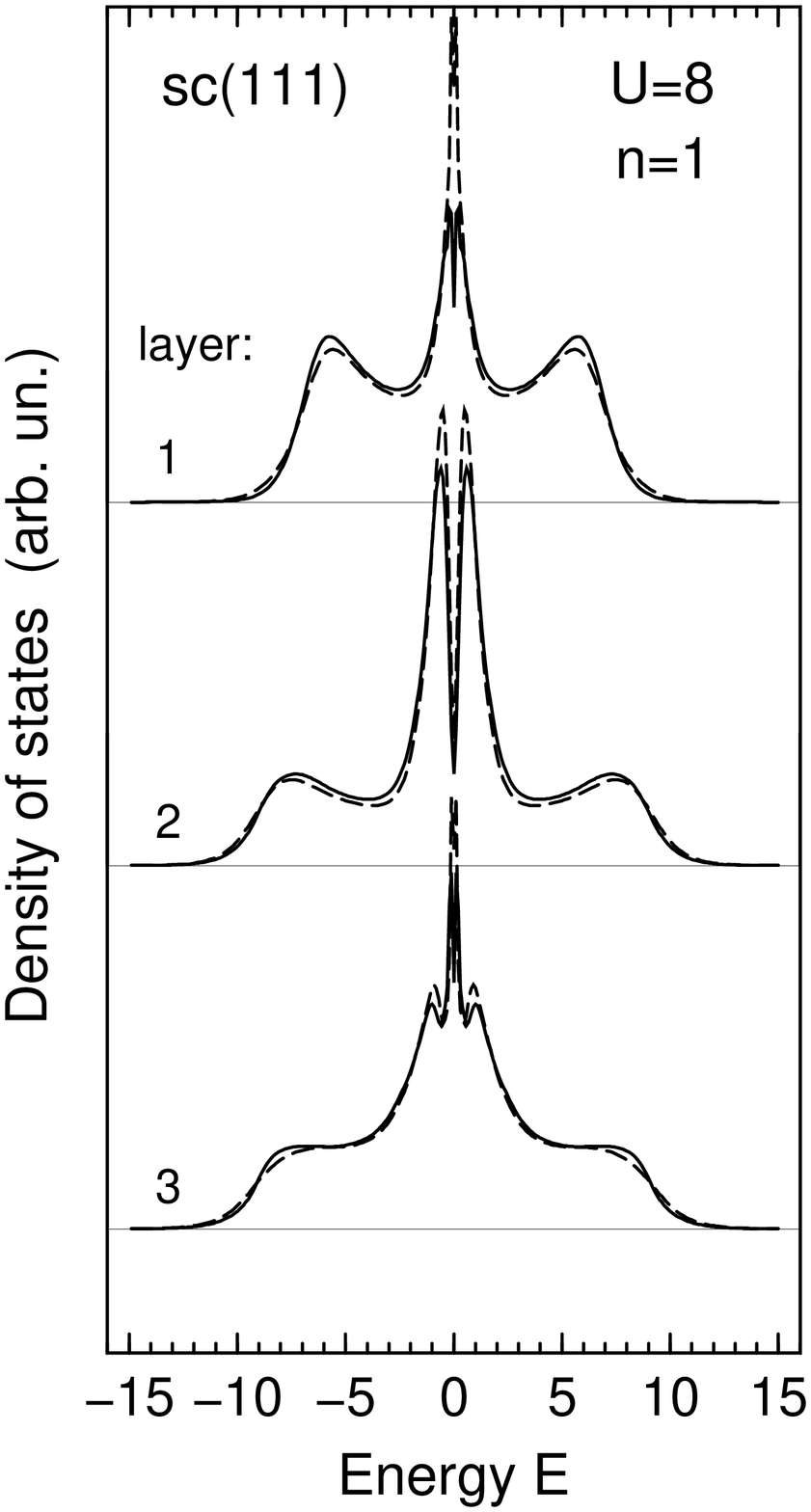}}
\end{picture}
\parbox[]{85mm}{\small Fig.~12.
The same as Fig.~11, but for the sc(111) surface.
}
\end{figure}

Finally let us investigate the effect of the non-local parts of the
self-energy on the QDOS. The local QDOS in the first three layers 
for $U=8$ and at half-filling ($n=1$) is shown in Fig.~11 for the 
(110) and in Fig.~12 for the (111) surface (solid lines). In both 
cases we find the spectra for all layers to consist of the two 
high-energy charge-excitation peaks as well as of the resonance 
at the Fermi level. Compared with the results for the (100) surface
(Fig.~8), the fine structure within the resonance is much stronger.
This has to be regarded as a pure surface effect since corresponding
structures are found in the free DOS at the respective surface 
as well. The narrowing of the top-layer QDOS is another surface 
effect that is more pronounced for the (110) and even more for 
the (111) than for the (100) surface. This is consistent with Eq.\ 
(\ref{eq:var}) when taking into account the coordination numbers of 
the top-layer sites at the respective surfaces. Due to the imaginary 
part of the self-energy the differences 
between the QDOS for different
layers are diminished to a large extent compared with $U=0$. If 
this can be extrapolated to the asymmetric case $n\ne 1$, at least
qualitatively, it implies a damping of the surface Friedel 
oscillations in the layer-dependence of the charge density.

The QDOS shown in Fig.~11 and Fig.~12 has been calculated including 
the non-local parts of the self-energy up the third neighbor shell
(solid lines). This is completely sufficient to obtain fully
converged results. For $s=1$, i.~e.\ including the on-site and the
nearest-neighbor non-local part only, we get almost the same 
results. It can thus be concluded that for all surfaces considered
the non-local terms beyond nearest neighbors are quite unimportant 
as concerning their effects on the QDOS. Differences to the fully 
converged results show up in the case $s=0$, i.~e.\ for the local 
approximation (dashed lines). However, even for the more open 
(110) and
(111) surfaces the local approximation qualitatively recovers all
features in the spectra. Concerning the top-layer QDOS, it is 
slightly worse than for the (100) surface. For the second- and the
third-layer QDOS the errors are comparable to those for the bulk
case. Again we find an overestimation of the weight of the 
resonance around $E=0$ while the weight of the charge-excitation
peaks is slightly too small.

{\center \bf \noindent IV. CONCLUSIONS \\ \mbox{} \\} 

The behavior of the electronic self-energy and of the 
quasi-particle density of states at the low-index surfaces 
of a simple cubic lattice has been investigated within the 
framework of the semi-infinite Hubbard model. Our main interest
has been focussed on the non-locality of the self-energy and
its effects on the electronic structure. Within the weak-coupling
approach that has been employed for the present
study, the non-locality 
can fully be taken into account. The results for infinitesimally
small interaction $U$ are exact. For finite $U$ and at half-filling
the SOPT-HF approach is expected to yield qualitatively reliable 
results at least. SOPT-HF studies have been performed previously
for infinitely extended lattices with full translational symmetry
(e.~g.\ Refs.\ \cite{SC91,MH89b}).

Let us briefly summarize the results found for the behavior of the
self-energy $\Sigma^{(2)}_{ij}(E)$ at a single-crystal surface: The 
reduced translational symmetry implies that there are inequivalent 
sites $i$ and $j$ within the same neighbor shell. This results in 
a directional dependence of the non-local self-energy parts. The 
directional dependence has been found to be weak and to be confined 
to the very surface. The local self-energy $\Sigma_{ii}^{(2)}(E)$ 
at the surface differs from its bulk value for the top layer only 
(in the case of the open (111) surface there are non-negligible 
differences also for the second layer). Differences of the non-local 
parts ($i\ne j$) compared with the bulk case are significant if 
there is at least one site within the top layer. As a consequence 
of the reduced coordination number of sites at the very surface, 
there is a narrowing of the free top-layer DOS. This results in 
a corresponding narrowing of the local top-layer self-energy. 
Furthermore, the reduced coordination number results in a 
diminished (enhanced) surface charge density below (above) 
half-filling which (in each case) implies a lowered total 
weight of the imaginary part of the top-layer self-energy.
The decrease of the absolute magnitudes of the non-local parts 
with increasing distance between the sites $i$ and $j$ is as 
fast as for the bulk, but considerable faster than for the 
two-dimensional case. All effects mentioned have been found to 
be the stronger the more open the surface.

The effects of the non-locality of the self-energy on the 
quasi-particle density of states can be summarized as follows:
Qualitatively, all effects that show up in the layer-dependent
QDOS at the surface are the same as those for the bulk QDOS:
The inclusion of more and more non-local self-energy parts 
reduces the weight of the quasi-particle resonance at the Fermi
level and pronounces the high-energy charge-excitation peaks.
Although the top-layer self-energy is sensitively dependent on
the type of the surface, the effects of its non-local parts on
the QDOS are of the same order of magnitude for the different 
surfaces considered. In all cases the local approximation 
is found to 
yield qualitatively correct results. The errors introduced
by the local approximation in the surface region are comparable to
those in the bulk. With respect to the QDOS the local approximation
is only slightly better than for the two-dimensional case. However, 
there is one main difference that distinguishes between the $D=2$ 
lattice on the one hand and the semi-infinite ($D=3$) lattice on 
the other: In the latter one gets nearly convergent results by 
taking into account merely the on-site and the nearest-neighbor 
self-energy parts. On the contrary, up to $s\approx5$ shells have 
to be included to obtain convergence for $D=2$.

It depends on the type of the physical problem to be investigated
whether or not one can tolerate the errors introduced by the
local approximation. This stresses the need for a quantitative
estimate of the effects of non-locality which has been given 
here for the semi-infinite Hubbard model. Our study should have 
shown that if one agrees on the use of the local approximation 
for the $D=3$ bulk, one may likewise consider a surface without 
the necessity for further justification. In particular, this is 
important for the application of the 
dynamical mean-field theory (DMFT) \cite{Vol93,GKKR96} to the 
semi-infinite Hubbard model. This means to apply those many-body 
techniques that have been developed for the (approximate or even
exact) solution of the infinite-dimensional Hubbard model. 
For $D=\infty$ the local approximation
becomes exact, and there is a universal functional relation
between the local self-energy and the QDOS. 
Within the context of the DMFT the main question thus concerns
the effects of the non-local parts of the self-energy on the 
QDOS which could be estimated within the present study.

{\center \bf \noindent ACKNOWLEDGEMENT \\ \mbox{} \\} 

Support of this work by the Deutsche Forschungsgemeinschaft within
the Sonderforschungsbereich 290 is gratefully acknowledged.
\vspace{5mm}

-------------------------------------------------------------------


\begin{thebibliography}{10}
\bibitem{tpdisc}
The overall three-peak structure of the QDOS has been observed
in previous SOPT-HF studies \cite{SC91,BJ90}. It is remarkable
that indications of the splitting into an upper and a lower
Hubbard band appear in a weak-coupling approach. This might
be ascribed to the fact that (at half-filling) the SOPT-HF
self-energy (accidentally) recovers the atomic limit and 
exhibits the correct high-energy behavior.
\end{thebibliography}
\end{document}